\documentclass[onecolumn,useAMS,usenatbib]{mn2e}

\usepackage[dvips,draft]{graphicx}

\title[Horizon-Penetrating Transonic Accretion Disks  
around Rotating Black Holes]{Horizon-Penetrating Transonic Accretion Disks  
around Rotating Black Holes}
\author[Rohta Takahashi]{Rohta Takahashi$^{1}$\thanks{E-mail:
rohta@ea.c.u-tokyo.ac.jp}\\
$^{1}$Graduate School of Arts and Sciences, University of Tokyo, Komaba, Meguro, 
Tokyo 153-8902, Japan}

\begin{document}

\date{Accepted 200X December 15. Received 200X December 14; 
in original form 200X October 11}

\pagerange{\pageref{firstpage}--\pageref{lastpage}} \pubyear{2007}

\maketitle

\label{firstpage}

\begin{abstract}
The stationary hydrodynamic equations for the transonic 
accretion disks and flows around rotating black holes 
are presented by using the Kerr-Schild coordinate    
where there is no coordinate singularity at the event horizon. 
We use two types of the causal viscosity prescription, 
and the boundary conditions 
for the transonic accretion flows are given at the sonic point. 
For one type of the causal viscosity prescription 
we also add the boundary conditions at the viscous point where the accreting 
radial velocity is nearly equal to the viscous diffusion velocity.  
Based on the formalism for the transonic accretion disks, 
after we present the calculation method of the transonic solutions,  
the horizon-penetrating transonic solutions which smoothly pass 
the event horizon are calculated for several types of 
the accretion flow models:  
the ideal isothermal flows, the ideal and the viscous polytropic flows,  
the advection dominated accretion flows (ADAFs) with 
the relativistic equation of state, 
the adiabatic accretion disks, the standard accretion disks, 
the supercritical accretion disks.   
These solutions are obtained for both non-rotating and rotating black holes. 
The calculated accretion flows plunge into black hole with finite 
three velocity 
smaller than the speed of light even at the event horizon or inside 
the horizon, 
and the angular velocities of the accretion flow at the horizon 
are generally 
different from the angular velocity of the frame-dragging due 
to the black hole's 
rotation. 
These features contrast to the results obtained by using the Boyer-Lindquist 
coordinate with the coordinate singularity at the horizon. 
\end{abstract}


\section{Introduction}
Elucidating the nature of the strong-gravity region around a black hole 
is one of the greatest challenges in astrophysics in this century. 
The most of the gravitational energy is converted into another 
forms of energies 
such as the kinetic, the thermal and/or the radiation energy, 
and especially the emission from the vicinity of the event horizon 
contains information 
about physics of the strong-gravity region such as the physical 
parameters of a black hole. 
Since such emission by e.g. photons or neutrinos is usually produced 
in the accretion flows 
plunging into the black hole, the structure of the accretion flows 
in the vicinity of 
the black hole's horizon is frequently required to be solved precisely. 
Especially, the effects of the frame dragging due 
to the black hole's rotation 
are taken into account only when 
the fully general relativistic calculations are done. 
In the past studies, 
for many astrophysical systems and situations, 
the stationary solutions for the structure of the transonic accretion flows 
near the horizon are solved for the standard accretion disks 
\citep{nt73,pt74}, 
the advection-dominated accretion flows 
\citep{c96,acgl97,jk97,gp98,pg98,m00}, 
the polytropic accretion flows \citep{pa97}, 
the super-critical accretion flows \citep{b98, sm03}, 
and the hypercritical accretion flow model for 
the neutrino-dominated accretion flow \citep{pwf99}. 
All these works use the Boyer-Lindquist coordinate 
where the coordinate singularity exists at the horizon. 
Due to the coordinate singularity, some physical values 
of the accretion flow 
based on the calculations using the Boyer-Lindquist coordinate 
are not realistic. 
For example, the radial component of three velocity 
of the accretion flow equals to the speed of light at the horizon, and 
the corresponding gamma factor diverges just outside the horizon.  
But these features are not the cases for the realistic accretion flow, i.e., 
the accretion flows plunge into the horizon with some finite velocity 
smaller than the speed of light, 
because from the point of view of the local observer moving 
along the fluid motion, 
the horizon is not a special location.   
So, one of the possible natural next step is to extend these past studies 
to the formulation using the Kerr-Schild coordinate where there 
is no coordinate singularity at the horizon. 
%
%
Our calculations in this study by using the Kerr-Schild coordinate 
avoiding the coordinate singularity show the radial velocity smaller 
than the speed of light at the horizon.  
In this study, we first show all the explicit formulation for the transonic 
accretion flows written by the Kerr-Schild coordinate, and then calculate 
the horizon-penetrating transonic solutions for the accretion flow 
which do not have any singularity at the horizon. 
The Kerr-Schild coordinates are frequently used in the past study, 
especially 
for the dynamical numerical calculations of the hydrodynamics 
or the magnetohydrodynamics around the black holes 
\citep{pf98, fip98, c00, f00, k01, gmt03, k04, gsm04}.  
While these studies mainly concentrate on the dynamical calculations of the 
accretion flows around the black hole, 
\cite{pf98} also give the simple solutions 
for the stationary accretion flows 
plunging into the black hole which exhibit the maximum value of the radial 
speed at the horizon. 
In our calculations, we found that this feature 
is not a general statement in the 
viscous accretion flows, i.e., the maximum value of the radial velocity 
is not generally achieved at the position of the horizon. 
In the present study, we concentrate on the stationary axisymmetric 
accretion flow in the equatorial plane. 
In this study, we use two types of the causal viscosity prescription. 
One is the simple treatment of the kinematic viscosity so as to vanish 
the shear stress at the horizon. 
The other is based on the description of the 
shear stress measured in the fluid's rest frame. 
The latter type of the causal viscosity is considered by \cite{ps94} 
for the Newtonian case and \cite{gp98} for the relativistic case. 
By using the latter types of the causal viscosity prescription, 
we do not need to put the boundary condition on the horizon, 
such as zero-boundary condition used in, e.g., \cite{nkh97}. 
In this study, we give the formulation with the effects of the heat flux 
(or the radiation cooling), the heat inertia and the relativistic enthalpy, 
and the relatively general forms of the equation of state.  
Based on this formalism, we give the sample transonic solutions 
for three types of 
accretion flow models: ideal isothermal disks, polytropic disks and  
advection-dominated accretion flows (ADAFs) 
with the relativistic equation of state, 
the adiabatic accretion disks, the standard accretion disks, and 
the supercritical accretion disks.     
We give the preliminaries for the calculations of the transonic solutions 
by using the Kerr-Schild coordinate in \S 2 containing such as 
the background metric and the frame and the frame transformation 
used in this study. 
In \S\ref{sec:BEs}, the basic equations are given; 
the mass conservation (\S\ref{sec:mass_cons}), 
the radial momentum equation (\S\ref{sec:rad_mom}), 
the angular momentum equation (\S\ref{sec:ang_mom}), 
the equation for vertical structure which is the momentum conservation 
in $\theta$-direction (\S\ref{sec:Htheta}), 
the thermodynamic equation and the energy equation (\S\ref{sec:EOS}), 
the turbulent shear stress based 
on the causal viscosity (\S\ref{sec:shear}). 
The boundary conditions for the viscous transonic solutions 
are given in \S\ref{sec:BCPAF};  
the boundary conditions at the sonic point (\S\ref{sec:SP}) 
and the viscous point (\S\ref{sec:VP}). 
In \S \ref{sec:coupled_eq}, we summarize 
the coupled differential equations to be solved. 
We give the calculation procedures in \S \ref{sec:calc_method}. 
Formulation and/or Numerical solutions 
for the horizon-penetrating solutions of the 
transonic accretion flows for different types of the accretion flow models 
are presented in \S\ref{sec:NSs}: 
the formulation and the numerical solutions for 
the ideal isothermal accretion flow (\S \ref{sec:IdIs}), 
the polytropic disks (\S \ref{sec:PD}), 
the ADAF with relativistic equation of state (\S \ref{sec:ADAF}),
the adiabatic accretion disk and 
the standard accretion disk (\S \ref{sec:AD}),
the formulation for the supercritical accretion disk (\S \ref{sec:SCD}). 
We give concluding remarks in the last section. 
While the basic structure of the basic equations 
for the transonic accretion flows are simple, 
the explicit expressions for some formula are lengthy. 
In order to clearly see the outline of the calculations,  
we put the details of lengthy formula in the Appendix and in the main body 
only the important formula are given. 
%

\section{Metric, Reference Frame and Velocity Fields}
\label{sec:preliminaries}
\subsection{Background Metric}
\label{sec:metric}
Throughout the present study, 
we assume the background geometry around the rotating black hole 
written by the Kerr-Schild coordinate described as 
\begin{equation}
ds^2=-\alpha^2 dt^2+\gamma_{ij}(dx^i+\beta^idt)(dx^j+\beta^jdt),
\end{equation}
%
where 
$i,~j=r,~\theta,~\phi$, and 
the nonzero components of the lapse function $\alpha$, 
the shift vector $\beta^i$ and the spatial matrix $\gamma_{ij}$ are 
given in the geometric units as 
%
\begin{eqnarray}
\alpha&=&\left(1+\frac{2mr}{\Sigma}\right)^{-1/2},~
\beta^r=\frac{2mr/\Sigma}{1+2mr/\Sigma},
\nonumber\\
\gamma_{rr}&=&1+\frac{2mr}{\Sigma},
\gamma_{\theta\theta}=\Sigma,~~
\gamma_{\phi\phi}=\frac{A\sin^2\theta}{\Sigma},
\gamma_{r\phi}=\gamma_{\phi r}
=-a\sin^2\theta\left(1+\frac{2mr}{\Sigma}\right).
\end{eqnarray}
%
Here, we use the geometric mass $m=GM/c^2$, 
$\Sigma=r^2+a^2\cos^2\theta$, $\Delta=r^2-2Mr+a^2$ and 
$A=(r^2+a^2)^2-a^2\Delta\sin^2\theta$, 
where $M$ is the black hole mass, $G$ is the gravitational constant and 
$c$ is the speed of light. 
Explicit forms of nonzero components of metric $g_{\mu\nu}$ 
and its inverse $g^{\mu\nu}$ are calculated in Appendix \ref{app:metric}. 
The position of the outer and inner horizon, $r_\pm$, 
is calculated from $\Delta=0$ 
as $r_\pm=m\pm(m^2-a^2)^{1/2}$.    
The angular velocity of the frame dragging due to 
the black hole's rotation is 
calculated as $\omega=-g_{t\phi}/g_{\phi\phi}=2mar/A$.  
Although some past studies use the metric in the equatorial plane, 
i.e. $\theta=\pi/2$, 
we basically formulate the basic equations by using 
the metric including $\theta$. 
This is because in the calculations of the vertical structure 
of the accretion disk 
which are performed in Sec. \ref{sec:Htheta} and 
Appendix \ref{app:ell*} need the metric including 
$\theta$, and we would like to have the consistency 
of the notation of the metric throughout the paper. 
But, we actually calculate the transonic solutions of the accretion 
flow in Sec.\ref{sec:NSs} by setting $\theta=\pi/2$. 
%

\subsection{Reference Frames}
\label{sec:frame}
In this study, we use three types of reference frames. 
The first is the Kerr-Schild coordinate frame (KSF), in which 
most of our calculations are done. 
The second is the fluid's rest frame (FRF), an orthonormal tetrad basis 
carried by observers moving along the fluid. 
The components of the four velocity measured in the FRF are described as 
\begin{equation}
u^{(t)}=-u_{(t)}=1,~~u^{(i)}=u_{(i)}=0,~~(i=r,~\theta,~\phi)
\end{equation}
where the bracket denote the physical quantities measured in the FRF. 
The third frames are calculated from a stationary congruences formed by 
observers with a future-directed unit vector orthogonal to 
$t=$constant whose components are given as 
\begin{equation}
u_{t}=-\alpha,~~
u_{i}=0,~~(i=r,~\theta,~\phi), 
\end{equation}
and 
\begin{equation}
u^{t}=\alpha^{-1},~~
u^{i}= -\alpha^{-1}\beta^i,~~(i=r,~\theta,~\phi),   
\end{equation}
respectively. 
For this congruences, since the vorticity tensor vanishes 
[e.g., A.10.2 in \cite{fn98}], 
this congruences is the congruences of locally non-rotating observers. 
So, this frame is usually called as the locally non-rotating reference frame 
(LNRF). 
By using the Boyer-Lindquist coordinate, this observer is moving with 
the angular velocity of the frame-dragging due to the black hole's rotation 
\citep{b70,bpt72}. 
On the other hand, 
by using the Kerr-Schild coordinate, 
since $\beta^r\neq 0$ and $\beta^\theta=\beta^\phi=0$, 
the observers with $u_\mu=-\alpha \delta^t_\mu$ is radially 
falling with $u^\theta=u^\phi=0$. 
For such observers, 
the nonzero components of the covariant and the contravariant 
four velocities are given as 
\begin{equation} 
u_t^{\rm LNRF}=-\alpha,~~
u^t_{\rm LNRF}=\frac{1}{\alpha},~~
u^r_{\rm LNRF}=-\frac{\beta^r}{\alpha}.    
\end{equation}
%
%
We can easily show that the LNRF is an orthonormal tetrad basis carried 
by the observer moving with $u^\mu_{\rm LNRF}$.  
The physical quantities measured in the LNRF are described by using the hat 
such as $u^{\hat{\mu}}$, $u_{\hat{\mu}}$, etc. 
In the Kerr-Schild coordinate, 
since the congruences of the observers moving with 
the angular velocity of the frame-dragging have the singularity 
at the event horizon as shown in the Appendix \ref{app:singular_frame}, 
we do not use such congruences in our calculations. 
%

\subsection{Frame Transformations and Velocity Fields}
\label{sec:frame_transformation}
The physical quantities measured in the KSF are transformed 
to those in the LNRF 
by tetrads $e_{\mu}^{~\hat{\nu}}$ and $e^{\mu}_{~\hat{\nu}}$. 
For example, the four velocity is transformed as 
$u^\mu=e^{\mu}_{~\hat{\nu}}u^{\hat{\nu}}$ and 
$u_\mu=e_{\mu}^{~\hat{\nu}}u_{\hat{\nu}}$. 
The explicit expressions of the tetrad components 
$e_{\mu}^{~\hat{\nu}}$ and $e^{\mu}_{~\hat{\nu}}$ are given 
in Appendix \ref{app:tetrads}.  
The FRF usually moves with some radial and azimuthal velocities with 
respect to the LNRF. 
We newly defined the radial velocity $v_{r}$ and 
the rotational velocity $\hat{v}_\phi$ such that 
the FRF moves with the radial velocity $v_{r}$ and 
the azimuthal velocity $\hat{v}_\phi$ with respect to the LNRF. 
By using these velocities, 
the physical quantities in the LNRF are transformed to those in the FRF 
by two-dimensional Lorentz transformation $e_{\hat{\nu}}^{~(\lambda)}$ 
and $e^{\hat{\nu}}_{~(\lambda)}$ with the radial velocity 
$v_{r}$ and the azimuthal velocity $\hat{v}_\phi$.  
Here, $e_{\hat{\nu}}^{~(\lambda)}$ and $e^{\hat{\nu}}_{~(\lambda)}$ are 
the transformation matrices denoting the two-dimensional Lorentz 
transformations, and the explicit expressions of these matrices  
are also given in Appendix \ref{app:tetrads}.  
By using the tetrads described in Appendix \ref{app:tetrads}, 
all the covariant and contravariant components 
of the four velocity in the KSF are calculated as
$u^\mu=e^\mu_{~\hat{\alpha}}e^{\hat{\alpha}}_{~(\nu)}u^{(\nu)}$ and 
$u_\mu=e_\mu^{~\hat{\alpha}}e_{\hat{\alpha}}^{~(\nu)}u_{(\nu)}$, 
and described by using 
the radial velocity $\hat{v}_r$ and the azimuthal velocity $\hat{v}_\phi$ 
of the FRF measured in the LNRF as shown in Appendix 
\ref{app:velocity_field}. 
Inversely, the radial velocity $\hat{v}_r$ and the 
azimuthal velocity $\hat{v}_\phi$ 
are described by the four velocities, $u^\mu$ and $u_\mu$, 
measured in the KSF 
as 
%
\begin{equation}
\hat{v}_r = \frac{u^r+\beta^r u^t}{\hat{\gamma}(\gamma_{rr})^{1/2}},~~~
\hat{v}_\phi = \frac{\ell}{\hat{\gamma}(\gamma_{\phi\phi})^{1/2}}, 
\label{eq:gam_vr_vphi}
\end{equation}
%
where $\hat{\gamma}\equiv(1-\hat{v}_r^2-\hat{v}_\phi^2)^{-1/2}=\alpha u^t$. 
Since $\hat{v}_r$ is the radial velocity measured in the LNRF which 
is radially falling with $u^{r}_{\rm LNRF}$, the radial velocity $\hat{v}_r$ 
can generally have both the positive and the negative values for the 
radially falling accretion flows. 
%
%


\subsection{$u^t$ and $\Omega$}
\label{sec:preliminaries}
When the transonic solutions are calculated later, 
we solve the differential equations for the radial four velocity $u^r$ 
and the angular momentum $\ell(=u_\phi)$. 
Therefore, it is convenient to express the angular velocity 
$\Omega$ and $u^t$ 
which are frequently used in the formula in the following sections 
by $u^r$ and $\ell$. 
From the normalization of the four velocity, $u^\mu u_\mu=-1$, we can obtain 
the quadratic equation of $u^t$ as 
$A_0 (u^t)^2+2B_0 u^t +C_0=0$ where 
$A_0= -\alpha^2+(\beta^r)^2/\gamma^{rr}$, 
$B_0= (\beta^r/\gamma^{rr})u^r$ and  
$C_0=(u^r)^2/\gamma^{rr}+\ell^2/\gamma_{\phi\phi}+1$. 
From the quadratic equation, we have the solution for $u^t>0$ as 
\begin{equation}
u^t=\frac{C_0}{D_0^{1/2}-B_0}, \label{eq:ut}
\end{equation}
where $D_0=B_0^2-A_0 C_0$. 
In this study, we consider the accretion flow with $u^r<0$. 
For such flows, $B_0<0$, and then $u^t>0$.  
We can also show that $D_0>0$ for the region $r>r_+$, 
and for the region $r<r_0$, 
$D_0>0$ only when $\ell^2<[(u^r)^2+1/\Sigma]A/(-\Delta)$.   
From $u^t$ calculated above, the angular velocity $\Omega$ is calculated 
as 
\begin{equation}
\Omega=\omega+\frac{\ell-\gamma_{r\phi}u^r}{u^t\gamma_{\phi\phi}} 
\label{eq:Om_ut}
\end{equation}
which is derived from $\ell=g_{\phi\mu}u^\mu$.  
Here, $u^t$ is calculated from Eq. (\ref{eq:ut}). 
We can also calculate $u_r$ and $\mathcal{E}(\equiv -u_t)$ 
from $u^r$ and $\ell$ as 
$u_r=(g_{tr}+g_{r\phi}\Omega)u^t+g_{rr}u^r$ and 
$\mathcal{E}=(g_{tt}+g_{t\phi}\Omega)u^t+g_{tr}u^r$ where 
$u^t$ and $\Omega$ are calculated by Eqs. (\ref{eq:ut}) 
and (\ref{eq:Om_ut}).  

\subsection{Transformation of Four Velocities written 
by Kerr-Schild Coordinate 
and Boyer-Lindquist Coordinate}
The transformation of the four velocities calculated 
by using the Boyer-Lindquist coordinate 
and the Kerr-Schild coordinate are given by 
%
\begin{eqnarray}
&&
u^t_{\rm BL}=u^t_{\rm KS}-\frac{2mr}{\Delta}u^r_{\rm KS},~~
u^r_{\rm BL}=u^r_{\rm KS},~~
u^\theta_{\rm BL}=u^\theta_{\rm KS},~~
u^\phi_{\rm BL}=u^\phi_{\rm KS}-\frac{a}{\Delta}u^r_{\rm KS}, 
\label{eq:uBLKS}\\
&&
u_t^{\rm BL}=u_t^{\rm KS},~~
u_r^{\rm BL}=u_r^{\rm KS}
+\frac{2mr}{\Delta}u_t^{\rm KS}+\frac{a}{\Delta}u_\phi^{\rm KS},~~
u_\theta^{\rm BL}=u_\theta^{\rm KS},~~
u_\phi^{\rm BL}=u_\phi^{\rm KS}.\label{eq:uBLKS2} 
\end{eqnarray}
%
Here, "BL" and "KS" denote the physical quantities calculated 
by using the Boyer-Lindquist 
coordinate and the Kerr-Schild coordinate, respectively. 
\footnote{
The transformation law given by Eq. (\ref{eq:uBLKS}) is calculated from 
the lapse function $\alpha$, the shift vector 
$\beta^\mu$ (or $\beta_\mu$) 
and the matrix $\gamma_{ij}$ 
(or $\gamma^{ij}$) for the Kerr-Schild coordinate as 
%
\begin{eqnarray}
u^t_{\rm BL}=u^t_{\rm KS}
	-\frac{\beta^r}{\alpha^2\gamma^{rr}-(\beta^r)^2}u^r_{\rm KS},~~~~~
u^\phi_{\rm BL}=u^\phi_{\rm KS}
	+\left(\frac{\gamma_{r\phi}}{\gamma_{\phi\phi}}\right)
	\frac{\alpha^2\gamma^{rr}}{\alpha^2\gamma^{rr}-(\beta^r)^2}
	u^r_{\rm KS},  \nonumber
\end{eqnarray}
%
which are derived from the metric expressed as 
%
\begin{eqnarray}
ds^2&=&
-\frac{\alpha^2\gamma^{rr}-(\beta^r)^2}{\gamma^{rr}}
	\left[
	dt-\frac{\beta^r}{\alpha^2\gamma^{rr}-(\beta^r)^2}dr
	\right]^2
+\left[\frac{\alpha^2}{\alpha^2\gamma^{rr}-(\beta^r)^2}\right] dr^2
+\gamma_{\theta\theta}d\theta^2
\nonumber\\
&&
+\gamma_{\phi\phi}
	\left\{
	d\phi +\left(\frac{\gamma_{r\phi}}{\gamma_{\phi\phi}}\right)
	\frac{\alpha^2\gamma^{rr}}{\alpha^2\gamma^{rr}-(\beta^r)^2}dr
	+\beta^r \left(\frac{\gamma_{r\phi}}{\gamma_{\phi\phi}}\right)
	\left[dt-\frac{\beta^r}{\alpha^2\gamma^{rr}-(\beta^r)^2}dr\right]
	\right\}^2 \nonumber
\end{eqnarray}
}

\section{Basic Equations}
\label{sec:BEs}
The basic equations for the relativistic hydrodynamics are 
the baryon-mass conservation $(\rho_0 u^\mu)_{;\mu}=0$ and 
the energy-momentum conservation $T^{\mu\nu}_{;\nu}=0$, where 
$\rho_0$ is the rest-mass density and $T^{\mu\nu}$ is the 
energy-momentum tensor. 
Dynamical basic equations except the baryon mass conservation 
are calculated from the energy-momentum tensor, $T^{\mu\nu}$. 
We use the energy-momentum tensor written as, 
%
\begin{equation}
T^{\mu\nu}=\rho_0\eta u^\mu u^\nu + pg^{\mu\nu} 
+ t^{\mu\nu}+q^\mu u^\nu+q^\nu u^\mu, 
\end{equation}
%
where 
$p$ is the pressure, 
$\eta=(\rho_0+u+p)/\rho_0$ is the relativistic enthalpy, 
$u$ is the internal energy,  
$t^{\mu\nu}$ is the viscous stress-energy tensor 
and 
$q^\mu$ is the heat-flux four vector.  
In the present study, we do not include the heat flux term in the 
energy-momentum tensor. 
One of the natural form of the shear stress, $t^{\mu\nu}$, is 
the Navier-Stokes shear stress. 
The relativistic Navier-Stokes shear stress, $t^{\mu\nu}$, is written as 
\citep{mtw73},  
%
\begin{equation}
t^{\mu\nu}=-2\lambda \sigma^{\mu\nu}-\zeta \Theta h^{\mu\nu},  
\end{equation}
%
where 
$\lambda$ is the coefficient of dynamic viscosity, 
$\xi$ is the coefficient of bulk viscosity, 
$h^{\mu\nu}\equiv g^{\mu\nu}+u^\mu u^\nu$ is the projection tensor, 
$\Theta\equiv u^\gamma_{~;\gamma}$ is the expansion of the fluid world line, 
and 
$\sigma^{\mu\nu}$ is the shear rate of the fluid which is calculated as 
%
\begin{eqnarray}
\sigma_{\mu\nu}
	&=&
	\frac{1}{2}(u_{\mu;\alpha}h^{\alpha}_\nu + u_{\nu;\alpha}h^{\alpha}_\mu)
	-\frac{1}{3}\Theta h_{\mu\nu}, \\
 	&=&
	\frac{1}{2}(u_{\mu;\nu}+u_{\nu;\mu}+a_\mu u_\nu + a_\nu u_\mu)
	-\frac{1}{3}\Theta h_{\mu\nu},
\end{eqnarray}
where $a_\mu\equiv u_{\mu;\gamma}u^\gamma$ is the four acceleration. 
In this study, we do not take the shear stress written by this form. 
Instead, we use the Kerr-Schild coordinate version of 
the shear stress used in \cite{gp98} and \cite{pg98} 
which allows angular momentum transport and preserve causality. 
We evaluate the shear stress in the FRF and assume that 
all the components of the shear stress except $t_{(r)(\phi)}=t_{(\phi)(r)}$ 
are null in the FRF. 
Based on this assumption, the shear stress measured in the KSF is calculated 
by using the tetrads connecting the KSF and the FRF, e.g.,  
$t^{\mu}_{~\nu}=2[e^{\mu(r)}e_\nu^{~(\phi)}
+e^{\mu(\phi)}e_\nu^{~(r)}]~t_{(r)(\phi)}$. 
The explicit forms of the shear stress in the FRF used 
in this study is given 
in Sec. \ref{sec:shear}. 
%

%
In this study, we consider the stationary, 
axisymmetric and equatorially symmetric 
global accretion flow in the equatorial plane, i.e., 
we assume $u_\theta=0$. 
We also assume that the effects of the bulk viscosity is negligible.    
In the following sections, we derive the basic equations 
written by the Kerr-Schild 
coordinate by using the vertical averaging procedures used 
in, e.g., \cite{gp98}, 
around the equatorial plane.  
%

\subsection{Mass Conservation and Mass-Energy Flux}
\label{sec:mass_cons}
The equation for the baryon mass conservation is written as 
%
\begin{equation}
(\rho_0 u^\mu)_{;\mu}=0, 
\end{equation}
%
where $\rho_0$ is the rest-mass density and $u^\mu$ is the four velocity. 
By averaging the physical quantities around the equatorial plane, 
the mass-accretion rate 
$\dot{M}$ is calculated as 
%
\begin{equation}
\dot{M}=-4\pi \sqrt{-g} H_\theta \rho_0 u^r, 
\label{eq:mass_cons}
\end{equation}
%
where $H_\theta$ is the half-thickness of the accretion flow 
in the $\theta$-direction 
which is calculated in Sec. \ref{sec:Htheta}, and $\sqrt{-g}=r^2$. 
When we calculate the global structure of the accretion flow, 
we normalize the rest-mass density, $\rho_0$, by setting $\dot{M}=1$, i.e., 
the mass conservation is written as 
\begin{equation}
-4\pi r^2 H_\theta \rho_0 u^r=1. 
\label{eq:mass_cons2}
\end{equation}
From the projection of the energy-momentum 
conservation, $T^{\mu\nu}_{~;\nu}=0$, 
onto $t$-component, i.e., $h^t_\mu T^{\mu\nu}_{~;\nu}=0$, with the vertical 
averaging calculations, we get
%
\begin{eqnarray}
\left[
	\eta\mathcal{E} +\frac{4\pi H_\theta \sqrt{-g}}{\dot{M}}
	(t^r_{~t}+u^rq_t-q^r\mathcal{E})
\right]_{,r}
=\frac{4\pi H_\theta \sqrt{-g}}{\dot{M}}\mathcal{E} 
q^\theta. \label{eq:diffEC}
\end{eqnarray}
%
From Eq. (\ref{eq:diffEC}), we obtain 
\begin{equation}
\eta\mathcal{E}-\epsilon_0+\frac{4\pi H_\theta 
\sqrt{-g}}{\dot{M}}t^r_{~t}=Q_{\mathcal{E}}, 
\end{equation}
where $\epsilon_0$ is the specific energy of the flow and 
$Q_{\mathcal{E}}$ represents the effects of the heat flux defined as 
\begin{equation}
Q_{\mathcal{E}}\equiv 
	\int_{r_{\rm min}}^{r_{\rm max}} 
		\frac{4\pi H_\theta \sqrt{-g}}{\dot{M}} \mathcal{E}q^\theta dr 
	-\frac{4\pi H_\theta \sqrt{-g}}{\dot{M}}
	\left(u^r q_t-q^r\mathcal{E}\right). 
\end{equation}
In the case of no heat flux, Eq. (\ref{eq:diffEC}) is reduced to  
%
\begin{eqnarray}
\dot{E}=\dot{M}\eta\mathcal{E}+4\pi r^2 H_\theta t^r_{~t},   
\end{eqnarray}
%
where $\dot{E}$ is the mass-energy flux corresponding to 
the rate of change of the black hole mass if measured at the horizon. 
With the mass conservation, we obtain the specific energy $\epsilon_0$ as 
%
\begin{eqnarray}
\epsilon_0 = \frac{\dot{E}}{\dot{M}}
=\eta\mathcal{E}+\frac{t^r_{~t}}{\rho_0 u^r}, 
\end{eqnarray}
%
where $\epsilon_0$ is the specific energy of the accreting matter. 
When the velocity of the accreting matter is non-relativistic 
and cold i.e. $\eta=1$ 
where the thermal energy of the matter is much lower than 
the rest-mass energy, the specific energy become unity, 
i.e. $\dot{E}\sim\dot{M}$. 
%
%

\subsection{Radial Momentum Conservation}
\label{sec:rad_mom}
The equation for the radial momentum conservation is obtained 
by the projection 
of the equation for the energy momentum conservation, 
$T^{\mu\nu}_{~;\nu}=0$, 
into $r$-direction, i.e., $h^r_\mu T^{\mu\nu}_{~;\nu}=0$.
We can write down $h^r_\mu T^{\mu\nu}_{~;\nu}=0$ as, 
%
\begin{equation}
a^r=-\frac{h^{rr}}{\rho_0\eta}\frac{dp}{dr}+n_{\rm HI}, 
\end{equation}
%
where $a^r$ is the radial component of the four-acceleration of the fluid, 
$a^\mu=u^\mu_{~;\nu}u^\nu$, and $n_{\rm HI}$ includes the effects of 
the heat inertia which are discussed by \cite{ban97}. 
The radial component of the four-acceleration, $a^r(=u^r_{~;\nu}u^\nu)$, 
is calculated as 
%
\begin{equation}
a^r=u^r\frac{du^r}{dr}-n_{\rm acc}, 
\end{equation}
where we decompose $n_{\rm acc}$ into three parts as 
\begin{equation}
n_{\rm acc}=n_{\rm KP}(\Omega-\Omega_K^+)(\Omega-\Omega_K^-)
+n_{\rm BL}+n_{\rm KS}.  
\end{equation}
%
Here, $\Omega\equiv u^\phi/u^t$ is the angular velocity and 
$\Omega_K^\pm$ are the Keplerian angular momentum described as 
$\Omega_K^\pm=\pm m^{1/2}/(r^{3/2}\pm am^{1/2})$ which are the solutions of 
$g_{\phi\phi,r}\Omega^2+2g_{t\phi,r}\Omega+g_{tt,r}=0$. 
The term including $n_{\rm KP}$ measures the deviation 
of the angular velocity 
from the Keplerian angular velocity.  
The terms $n_{\rm KP}$, $n_{\rm BL}$ and $n_{\rm KS}$ are given as 
%
\begin{eqnarray}
n_{\rm KP} &\equiv& \frac{1}{2}g^{rr}g_{\phi\phi,r}(u^t)^2,\\ 
n_{\rm BL}&\equiv& -\frac{1}{2}g^{rr}g_{rr,r}(u^r)^2,\\ 
n_{\rm KS}&\equiv&-(g^{tr}g_{tr,r}+g^{r\phi}g_{r\phi,r})(u^r)^2 
	-\left[
		g^{tr}\left(g_{tt,r}+g_{t\phi,r}\Omega\right)
		+g^{r\phi}\left(g_{t\phi,r}+g_{\phi\phi,r}\Omega\right)
	\right]u^tu^r . \nonumber\\
\end{eqnarray}
%
Since $n_{\rm KS}$ contains $g^{tr}$ and $g^{r\phi}$ which are null 
for the Kerr metric written by the Boyer-Lindquist coordinate, 
this term is newly calculated term in this study which use the Kerr-Schild 
coordinate. 
On the other hand, the general form of $n_{\rm HI}$ is described as 
%
\begin{eqnarray}
n_{\rm HI}&\equiv& -\frac{1}{\rho_0\eta}
	\left[ 
		h^r_\mu t^{\mu\nu}_{~;\nu}+h^r_\mu (q^\mu u^\nu+q^\nu u^\mu)_{;\nu} 
	\right]\label{eq:nHIdefinition}\\
&=& -\frac{1}{\rho_0\eta}
	\bigg[
		\left( 
			\Phi -\frac{1}{3}\Theta t^\gamma_{~\gamma}
			+u_\mu u^\nu q^\mu_{~;\nu}
		\right)u^r
		+t^{r\nu}_{~;\nu}
		+q^r_{;\gamma}u^\gamma +q^r\Theta +u^r_{;\nu}q^\nu
	\bigg], 
\label{eq:nHI_b}
\end{eqnarray}
where 
$\Phi\equiv -\sigma_{\mu\nu}t^{\mu\nu}$ is the dissipation function 
which is calculated in Sec. \ref{sec:shear} based on 
the shear stress measured in the FRF, and  
$-(1/3)\Theta t^\gamma_{~\gamma}$ represents the compressive heating rate.  
If the effects of the dissipation function, $\Phi$, 
is dominated, $n_{\rm HI}$ is reduced to 
\begin{equation}
n_{\rm HI}=-\frac{u^r}{\rho_0\eta}\Phi, 
\label{eq:n_HeatInertia}
\end{equation}
and we use this expression in this study. 
Finally, 
from the equation for the radial momentum conservation, 
we can derive the equation for $du^r/dr$ as, 
%
\begin{equation}
u^r \frac{du^r}{dr} = 
-\frac{h^{rr}}{\rho_0\eta}\frac{dp}{dr}+n_{\rm acc}+n_{\rm HI}.   
\label{eq:radmom}
\end{equation}
%
This equation is used to derive equations which determine 
the boundary conditions 
for the sonic point and the viscous point 
in Sec. \ref{sec:BCPAF}. 
In this study, we solve the radial component of the four 
velocity $u^r$ in the KSR 
instead of the radial velocity $\hat{v}_r$ measured in the LNRF 
when we solve the transonic solutions. 
This is because while 
$\hat{v}_r$ have the negative and positive values as denoted above, 
$u^r$ is always negative for the accretion flow. 
Thus, we choose $u^r$ as one of the basic dynamic variables 
when solving the transonic 
flows. 
%

\subsection{Angular Momentum Conservation}
\label{sec:ang_mom}
The equation for the angular momentum conservation is 
obtained by the projection 
of the equation for the energy momentum conservation, 
$T^{\mu\nu}_{~;\nu}=0$, 
into $\phi$-direction, i.e., $h^\phi_\mu T^{\mu\nu}_{~;\nu}=0$.
By using the vertical averaging procedure, 
we can write down $h^\phi_\mu T^{\mu\nu}_{~;\nu}=0$ as, 
%
\begin{eqnarray}
\left[
	\eta\ell -\frac{4\pi H_\theta \sqrt{-g}}{\dot{M}}
	(t^r_{~\phi}+u^rq_\phi-q^r\ell)
\right]_{,r}
=\frac{4\pi H_\theta \sqrt{-g}}{\dot{M}}\ell q^\theta. \label{eq:diffAMC}
\end{eqnarray}
%
From Eq. (\ref{eq:diffAMC}), we obtain 
\begin{equation}
\eta\ell-j-\frac{4\pi H_\theta \sqrt{-g}}{\dot{M}}t^r_{~\phi}=Q_{\ell}, 
\end{equation}
where $j$ is the specific angular momentum and 
$Q_{\ell}$ represents the effects of the heat flux defined as 
\begin{equation}
Q_{\ell}\equiv 
	\int_{r_{\rm min}}^{r_{\rm max}} 
		\frac{4\pi H_\theta \sqrt{-g}}{\dot{M}} \ell q^\theta dr 
	+\frac{4\pi H_\theta \sqrt{-g}}{\dot{M}}\left(u^r q_\phi-q^r\ell\right). 
\end{equation}
When the angular momentum is not transported by the heat flux, 
%
\begin{equation}
\dot{M}j=
	\dot{M}\eta\ell
	-4\pi r^2 H_\theta  t^r_{~\phi}, 
\label{eq:angmom}
\end{equation}
%
where 
the first term in the left-side hand is the total flux of 
the angular momentum of the fluid, 
the second term represents the amount of the dissipation due 
to the shear stress, 
and the right-hand side is the total inward flux of the angular momentum. 
In this study, we assume no angular momentum is transported by the heat flux, and in this case 
the shear stress tensor $t^r_{~\phi}$ is written as 
\begin{equation}
t^r_{\rm \phi}=-\rho_0 u^r (\eta \ell-j). 
\label{eq:angmom2}
\end{equation}
%

\subsection{Vertical Structure}
\label{sec:Htheta}
The equation for the vertical structure is calculated from the equation of 
the momentum conservation in $\theta$-direction by assuming the hydrostatic 
equilibrium. 
The equation for the momentum conservation in $\theta$-direction 
is obtained by the projection of the equation for the 
energy momentum conservation, $T^{\mu\nu}_{~;\nu}=0$, 
into $\theta$-direction, i.e., $h^\theta_\mu T^{\mu\nu}_{~;\nu}=0$.
Although 
the calculation procedures for the characteristic angular scale of the 
accretion flow, $H_\theta$, is basically same as those used in 
\cite{alp97}, we use several different assumptions. 
From $h^\theta_\mu T^{\mu\nu}_{~;\nu}=0$, 
by neglecting the effects due to the heat flux, i.e. $q^\mu=0$, 
we obtain 
%
\begin{eqnarray}
\frac{1}{\rho_0\eta}\frac{\partial p}{\partial \theta}
	&=&-\left( 
		u^r\frac{\partial u_\theta}{\partial r}
		+u^\theta\frac{\partial u_\theta}{\partial \theta}
	\right)
	+\Gamma^\mu_{\theta\nu}u_\mu u^\nu 
	-\frac{u_\theta}{\eta}
		\left(
			u^r\frac{\partial\eta}{\partial r}
			+u^\theta\frac{\partial\eta}{\partial\theta} 
		\right)
	-\frac{t^\nu_{~\theta;\nu}}{\rho_0\eta}
	-\frac{(q^\nu u_\theta+q_\theta u^\nu)_{;\nu}}{\rho_0\eta}.
\end{eqnarray}
%
Here, the terms including the differentiation of $\eta$ 
and $t^\nu_{~\theta;\nu}$ 
are newly considered terms which are not taking into account 
in \cite{alp97} which 
assume $\eta=1$.  
It is note that \cite{pg98} which do not assume $\eta=1$ 
use the equation for $H_\theta$ derived by \cite{alp97} assuming $\eta=1$. 
We expand the pressure in $\theta$-direction until 
the order of $\cos^2\theta$ as, 
%
\begin{equation}
p(r,\theta)=p_0(r)\left[1-\frac{1}{2}
\left(\frac{\cos\theta}{H_\theta}\right)^2\right],   
\end{equation}
%
where $p_0(r)$ is the pressure in the equatorial plane. 
This expansion is different from the expansion in \cite{alp97} 
by the factor $1/2$ before 
$(\cos\theta/H_\theta)^2$.  
From this, we can calculate 
$\partial p/\partial \theta=(2\cos\theta/H_\theta^2)p_0(r)$. 
From this, the angular half-thickness of the disk is calculated as 
%
\begin{eqnarray}
H_\theta^2
=\left(\frac{p_0}{\rho_0\eta}\right)\bigg/
	\bigg[
	-u^\nu u_{\theta,\nu}
	+\Gamma^\mu_{\theta\nu}u_\mu u^\nu
	-\frac{u_\theta u^\nu \eta_{,\nu}}{\eta}
	-\frac{t^\nu_{~\theta;\nu}}{\rho_0\eta}
	-\frac{(q^\nu u_\theta+q_\theta u^\nu)_{;\nu}}{\rho_0\eta}
	\bigg]
\label{eq:HtA}
\end{eqnarray}
%
In order to expand the denominator of Eq. (\ref{eq:HtA})
until the order of $\cos\theta$, 
we also approximate $u_\theta$ as 
\begin{equation}
u_\theta(r,\theta)=u_{\theta 1}(r)\cos\theta, 
\end{equation}
where $u_{\theta 1}$ is $\theta$-component of $u_\mu$ 
in the equatorial plane. 
By using these expansion until the order of $\cos\theta$, 
we obtain
%
\begin{eqnarray}
-u^\nu u_{\theta,\nu}&=&
	-\left( 
		u^r\frac{\partial u_\theta}{\partial r}
		+u^\theta\frac{\partial u_\theta}{\partial \theta}
	\right) \nonumber \\
	&=&\cos\theta \left(
		-u^r u_{\theta 1,r}+g^{\theta\theta}u_{\theta 1}^2
	\right), 
	\label{eq:ell*term1}\\
-\frac{u_\theta u^\nu\eta_{,\nu}}{\eta}&=&
	-\frac{u_\theta}{\eta}
		\left(
			u^r\frac{\partial\eta}{\partial r}
			+u^\theta\frac{\partial\eta}{\partial\theta} 
		\right) \nonumber \\
	&=&\cos\theta\left(
		-u_{\theta 1}u^r\frac{\partial \ln\eta}{\partial r}
	\right). \label{eq:ell*term3}
\end{eqnarray}
%
In the same way, 
we can also calculate $\Gamma^\mu_{\theta\nu}u_\mu u^\nu$ and 
$-t^\nu_{~\theta;\nu}/(\rho_0\eta)$ until the order of $\cos\theta$ as, 
\begin{eqnarray}
\Gamma^\mu_{\theta\nu}u_\mu u^\nu &=& 
\cos\theta~(\Gamma^\mu_{\theta\nu}u_\mu u^\nu)_1, \\
t^\nu_{~\theta;\nu}/(\rho_0\eta) &=& 
\cos\theta~[t^\nu_{~\theta;\nu}/(\rho_0\eta)]_1, \\
(q^\nu u_\theta+q_\theta u^\nu)_{;\nu}/(\rho_0\eta)
&=&\cos\theta~[(q^\nu u_\theta+q_\theta u^\nu)_{;\nu}/(\rho_0\eta)]_1, 
\end{eqnarray}
where the coefficients of terms of the order of $\cos\theta$ are defined as 
$(\Gamma^\mu_{\theta\nu}u_\mu u^\nu)_1$, 
$[t^\nu_{~\theta;\nu}/(\rho_0\eta)]_1$ and 
$[(q^\nu u_\theta+q_\theta u^\nu)_{;\nu}/(\rho_0\eta)]_1$. 
Direct calculations leads the explicit forms 
of $(\Gamma^\mu_{\theta\nu}u_\mu u^\nu)_1$ as 
\begin{equation}
(\Gamma^\mu_{\theta\nu}u_\mu u^\nu)_1=\ell_*^2/r^2, 
\end{equation}
where $\ell_*^2$ is calculated as
\begin{equation}
\ell_*^2 \equiv \ell^2-a^2(\mathcal{E}^2-1). 
\label{eq:ell*}
\end{equation}
From Eq. (\ref{eq:ell*}), $\ell_*=\ell$ when $a=0$ or $\mathcal{E}=1$. 
The form of Eq. (\ref{eq:ell*}) is same as the results of \cite{alp97} 
where the Boyer-Lindquist coordinate is used. 
This is because the transformations for $\ell$ and $\mathcal{E}$ between 
the Boyer-Lindquist coordinate and the Kerr-Schild coordinate are given as 
$\ell_{\rm BL}=\ell_{\rm KS}$ and 
$\mathcal{E}_{\rm BL}=\mathcal{E}_{\rm KS}$. 
In Appendix \ref{app:ell*}, we also show the direct derivations 
of Eq. (\ref{eq:ell*}) which is essentially same as the calculations 
using the Boyer-Lindquist coordinate as shown 
in \cite{alp97}, but several points are different.  
Then, the most general form for $H_\theta^2$ is calculated as 
%
\begin{eqnarray}
H_\theta^2
=\left(\frac{p_0}{\rho_0\eta}\right)\bigg/
	\bigg\{
		\frac{\ell_*^2}{r^2}
		-u^r u_{\theta,r}
		+u_{\theta} \left(u^{\theta} 
		-u^r \frac{\partial \ln\eta}{\partial r} \right) 
	-\left[\frac{t^\nu_{~\theta;\nu}}{\rho_0\eta} \right]_1
	-\left[\frac{(q^\nu u_\theta+q_\theta u^\nu)_{;\nu}}{\rho_0\eta}\right]_1
	\bigg\},
\label{eq:HtA}
\end{eqnarray}
%
where all physical values such as $c_s$, $u^r$, $u^\theta$ etc. 
are evaluated 
at the equatorial plane. 
In the present study, we assume $u_\theta=u_{\theta,r}=0$ and 
the negligible effects for the viscosity and the heat flux. 
These assumptions are basically same as \cite{acgl97} and \cite{gp98}. 
For the calculations of the transonic solutions in the later sections, 
we use the angular half-thickness of the disk described as, 
%
\begin{equation}
H_\theta=\frac{c_s}{\ell_*/r}.  
\label{eq:Ht2KS}
\end{equation}

\subsection{Energy Equation}
\label{sec:EOS}
The equation for the local energy conservation 
is obtained from $u_\mu T^{\mu\nu}_{~;\nu}=0$ as 
%
\begin{equation}
u^r\left(\frac{du}{dr}-\frac{u+p}{\rho_0}\frac{d\rho_0}{dr}\right)
=q_{\rm vis}^+ - q_{\rm rad}^-,   
\label{eq:energy}
\end{equation}
%
where 
\begin{eqnarray}
q_{\rm vis}^+&=&\Phi-\frac{1}{3}\Theta t^\gamma_{~\gamma},\\
q_{\rm rad}^-&=&-q^\mu_{~;\mu}-q^\mu a_\mu. 
\end{eqnarray}
Here the dissipation function $\Phi$ is given in Sec. \ref{sec:shear} and 
the second term in the right hand side of $q_{\rm vis}$ side represents 
the compressive heating rate. 
On the other hand, in the right hand side of $q_{\rm rad}^-$, 
$-q^\mu_{~;\mu}$ is the mass-energy flux transported out (in) 
to (from) the outside region, 
and $-q^\mu a_\mu$ is the special relativistic correction 
to $-q^\mu_{~;\mu}$ due to 
the heat inertia of the flux and represents the effects 
of the redshift of the flux. 
Since the left-hand-side of Eq. (\ref{eq:energy}) include the 
change of the entropy, $s$, as 
%
\begin{equation} 
\rho_0 T u^r \frac{ds}{dr}=q_{\rm vis}^+ - q_{\rm rad}^-,
\label{eq:energy2}
\end{equation}
%
where $\rho_0 T ds/dr$ represents the advected energy of the accretion flow, 
this equation represents the energy balance of the accretion flows, i.e. 
(advection cooling)=(viscous heating)-(radiative cooling). 
For the isothermal flows or the polytropic flows calculated 
in the later sections, 
we do not use the energy equation given by Eq. (\ref{eq:energy}) when we 
solve the transonic solutions for these flows. 
On the other hand, for the general equation of state where 
the pressure and the internal energy usually are the functions of both 
the rest-mass density $\rho_0$ and the temperature $T$, 
the energy equation given by Eq. (\ref{eq:energy}) is required in order to 
solve the transonic solution. 
As an example of such cases, we solve the transonic solutions for the 
advection dominated accretion flows with the general relativistic 
equation of state. 
%

\subsection{Treatment of Viscosity limited by Causality}
\label{sec:shear}
The viscosity due to the turbulent motion of magnetic field, 
fluids and particles 
such as photons and neutrinos is usually considered in the accretion disk. 
The effect of viscosity is transported to the finite length with 
the finite viscous timescale, $\tau_v$, and 
this transportation is limited by the causality. 
When the fluid's velocity approach the speed of light as near the horizon, 
it is expected that the viscous transportation become less effective. 
In this study, we phenomenologically take into account 
the causal viscous effects. 
The valid treatment of the causal viscosity will be required 
in future studies.
%
\subsubsection{Type A Causal Viscosity : Simple treatment of 
kinematic viscosity}
Here, we consider the kinematic viscosity by taking into account the 
causality. 
We use the kinematic viscosity coefficient, $\nu$, in order that 
the kinematic viscosity vanish on and inside the horizon which 
is expressed as 
\begin{equation}
\nu=
\left\{
	\begin{array}{ll}
	\nu_0 f_c, & {\rm for}~~~~r_+<r, \\
	0 & {\rm for}~~~~r\lid r_+, 
	\end{array}
\right.
\end{equation}
where $r_+$ is the radius of the horizon and 
$\nu_0$ is the kinematic viscosity coefficient when the effects of 
the causality is not considered, and $f_c$ is a cut-off function 
described as \citep{n92,pa97}, 
\begin{equation}
f_c=
\left\{
	\begin{array}{ll}
	[1-(\hat{v}/c_v)^2]^2, & {\rm for}~~~~|\hat{v}|\lid c_v, \\
	0, & {\rm for}~~~~|\hat{v}|>c_v.
	\end{array}
\right.
\end{equation}
Here, $\hat{v}=(1-\hat{\gamma}^{-2})^{1/2}$ where 
$\hat{\gamma}=\alpha u^t$. 
These treatments of the kinematic viscosity is similar to 
those of \cite{pa97}, 
but several points and explicit expression are different. 
By using the kinematic viscosity coefficient defined above, we calculate the 
shear viscosity tensor $t_{\mu\nu}$ as Navier-Stokes viscosity described as 
$t_{\mu\nu}=-2\rho_0\eta\nu\sigma_{\mu\nu}$. 
%

\subsubsection{Type B Causal Viscosity : Shear stress measured 
in fluid's rest frame}
%
In relativity, the physical meanings are not usually expressed directly 
in arbitrary frames. 
The FRF is the most natural place to evaluate the physical processes. 
We calculate the shear stress in the FRF as \cite{gp98} 
by using the relativistic version for the causal stress prescription 
proposed by \cite{ps94}. 
We assume the shear stress $t_{(r)(\phi)}=t_{(\phi)(r)}(\equiv S)$ 
in the FRF.  
The other components of the shear stress in the FRF 
except $t_{(r)(\phi)}$ and $t_{(\phi)(r)}$ are assumed to be null. 
This treatment is same as \cite{gp98}. 
By using the tetrads connecting the KSF and the FRF, 
the shear stress $t^r_{~\phi}$ is calculated as, 
%
\begin{equation}
t^r_{~\phi}=
	2 \left[
		e^{r(r)}e_{\phi}^{~(\phi)}+e^{r(\phi)}e_{\phi}^{~(r)}
	\right]
	t_{(r)(\phi)}
	=FS, 
\label{eq:tr_phi}
\end{equation}
%
where $F\equiv e^{r(r)}e_{\phi}^{~(\phi)}+e^{r(\phi)}e_{\phi}^{~(r)}$ and 
tetrads in $F$ are calculated by using the LNRF as 
$e^{\alpha(\beta)}=e^{\alpha}_{~(\beta)}
=e^\alpha_{~\hat{\mu}}e^{\hat{\mu}}_{~(\beta)}$ 
($\alpha= r$, $\beta= r,~\phi$)and 
$e_\alpha^{~(\beta)}=e_\alpha^{~\hat{\mu}}e_{\hat{\mu}}^{~(\beta)}$ 
($\alpha=\phi$, $\beta= r,~\phi$). 
The explicit forms of tetrad components are given 
in Appendix \ref{app:tetrads}. 
%
%
For the finite value of $t_{(r)(\phi)}$, the shear stress measured 
in the KSF, 
$t^r_{~\phi}$, is not null. 
This feature is contrasted to the shear stress calculated by using 
the Boyer-Lindquist coordinate as \cite{gp98}, see Eq. (60) in \cite{gp98}. 
The equation for the shear stress $S$ is described as \citep{gp98} 
%
\begin{equation}
u^r\frac{dS}{dr}=-\frac{S-S_0}{\tau_v}, 
\label{eq:dSdr}
\end{equation}
%
where $\tau_v$ is the relaxation timescale of the viscous diffusion and 
$S_0$ is the equilibrium value of the shear stress. 
The relaxation timescale $\tau_v$ is 
related to the propagation speed of the viscous effects $c_v$ as 
$c_v=(\nu/\tau_v)^{1/2}$ where $\nu$ is the kinematic viscosity. 
The coefficient of the dynamic viscosity $\lambda$ is described by the 
kinematic viscosity $\nu$ as $\lambda=\rho_0 \eta \nu$. 
In this study, the relaxation timescale $\tau_v$ is assumed 
as $\tau_v=1/\Omega$.
From these relations, the propagation speed of the viscous effects, 
$c_v$, are described by the sound speed, $c_s$, as 
$c_v=\alpha^{1/2}c_s$. 
These treatments are basically same as \cite{gp98}. 
From the angular momentum equation (\ref{eq:angmom2}), the shear stress $S$ 
is calculated as $S=-\rho_0 u^r(\eta\ell-j-Q_\ell)/F$. 
By differentiating this equation by $r$, we obtain $dS/dr$ and substitute 
$dS/dr$ to Eq. (\ref{eq:dSdr}). 
Then, the shear stress $S$ is calculated as, 
%
\begin{equation}
S=\frac{S_0+{\strut{\frac{\displaystyle\rho_0\eta\tau_v(u^r)^2}{
\displaystyle F}}}
		\left(
			\strut{ 
				\displaystyle
				\frac{d\ell}{dr}
				+\ell\frac{d\ln\eta}{dr}-\frac{1}{\eta}\frac{dQ_\ell}{dr}
			}
		\right)
	}
	{1-u^r\tau_v
		\left(
			\strut{
				\displaystyle
				\frac{2}{r}+\frac{d\ln F}{dr}+\frac{d\ln H_\theta}{dr}
			}
		\right)}.   
\label{eq:S}
\end{equation}
%
where $dQ_\ell/dr$ is determined by the heat flux and defined as 
\begin{equation}
\frac{dQ_\ell}{dr}=
	\frac{4\pi H_\theta\sqrt{-g}}{\dot{M}}\ell q^\theta
	+\left[\frac{4\pi H_\theta \sqrt{-g}}{\dot{M}}
		\left(u^rq_\phi-q^r \ell\right)\right]_{,r}.  
\end{equation}
From Eq. (\ref{eq:diffAMC}), $dQ_\ell/dr$ is also calculated as  
\begin{equation}
\frac{dQ_\ell}{dr}=\left[\eta\ell
	-\frac{4\pi H_\theta \sqrt{-g}}{\dot{M}}t^r_{~\phi}\right]_{,r}. 
\end{equation}
The equilibrium value of the shear stress $S_0$ is assumed to be 
the Navier-Stokes value as 
%
\begin{equation}
S_0=-2\rho_0\eta\nu\sigma_{(r)(\phi)}. 
\label{eq:S0}
\end{equation}
%
The shear rate $\sigma_{(r)(\phi)}$ in the FRF is calculated by using the 
shear tensor $\sigma_{\mu\nu}$ in KSF as 
$\sigma_{(r)(\phi)}=\sigma_{\mu\nu}~e^{\mu}_{~(r)}e^{\nu}_{~(\phi)}$. 
In Appendix \ref{app:shear}, 
we give the explicit forms for the shear tensors $\sigma_{\mu\nu}$ 
and the final form 
of $\sigma_{(r)(\phi)}$. 
These calculations are more lengthy than the same calculations 
using the Boyer-Lindquist 
coordinate but straightforward. 
Here, we simply express the shear rate $\sigma_{(r)(\phi)}$ as 
%
\begin{equation}
\sigma_{(r)(\phi)}\equiv\sigma
	=\sigma_r
	+\sigma_{u}\frac{du^r}{dr}
	+\sigma_\ell\frac{d\ell}{dr}.
\label{eq:shear0} 
\end{equation}
%
From Eqs. (\ref{eq:S}), (\ref{eq:S0}) and (\ref{eq:shear0}), we can derive 
the equation for $d\ell/dr$ having the singular point 
which we call viscous point. 
%

%
Based on the shear stress and the shear rate calculated in the last section, 
the dissipation function $\Phi$ is calculated as 
$\Phi\equiv-\sigma_{\mu\nu}t^{\mu\nu} 
	=-\sigma_{(\alpha)(\beta)}t^{(\alpha)(\beta)}
	=-2\sigma_{(r)(\phi)}t^{(r)(\phi)}$, 
and finally we obtain 
%
\begin{equation}
\Phi=-2\sigma S,  
\label{eq:diss_func}
\end{equation}
%
where $\sigma$ and $S$ are given by Eqs. (\ref{eq:S}) and (\ref{eq:shear0}).

\section{Boundary Conditions}
\label{sec:BCPAF}
The accretion flows plunging into the black hole supersonically 
must pass the sonic point where the accretion velocity become 
larger than the sound speed. 
On the other hand, when the causal viscosity prescription is used, 
the accretion flows pass the viscous point 
where the accretion velocity become 
larger than the speed of the viscous diffusion. 
In order to smoothly pass the sonic point and the viscous point, 
the flows must satisfy the boundary conditions 
at the sonic point (\S \ref{sec:SP}) 
and the viscous point (\S \ref{sec:VP}). 
By using the causal viscous prescription, 
the boundary conditions at the outer regions or the inner regions 
of the accretion flows
are not required. 
The boundary conditions at the event horizon which are used 
in some past studies 
are not required in the present study. 
%

\subsection{Boundary Conditions at the Sonic Point}
\label{sec:SP}
In order to obtain the boundary conditions at the sonic point, 
we need the equation which do not contain the derivatives except 
$du^r/dr$. 
The pressure $p$ and the internal energy $u$ 
of the accreting fluid is usually a function of the rest-mass 
density $\rho_0$ and/or the temperature $T$, i.e. $p=p(\rho_0,T)$ and 
$u=u(\rho_0,T)$.  
Then, the derivative $dp/dr$ is calculated as 
\begin{eqnarray}
\frac{dp}{dr}&=&
\left(\frac{\partial p}{\partial \rho_0}\right)_T \frac{d\rho_0}{dr}
+\left(\frac{\partial p}{\partial T}\right)_{\rho_0} \frac{dT}{dr},
\label{eq:dp_rho0T}\\ 
\frac{du}{dr}&=&
\left(\frac{\partial u}{\partial \rho_0}\right)_T \frac{d\rho_0}{dr}
+\left(\frac{\partial u}{\partial T}\right)_{\rho_0} \frac{dT}{dr}. 
\label{eq:du_rho0T}
\end{eqnarray}
In the case of the pressure and the internal energy is a function of 
the rest-mass density only, all the thermodynamic quantities can be 
written by the rest-mass density only, i.e. 
$(\partial p/\partial T)_{\rho_0}=(\partial u/\partial T)_{\rho_0}=0$.  
In such case, with Eqs. (\ref{eq:dp_rho0T}), (\ref{eq:du_rho0T}) and 
the mass conservation given by Eq. (\ref{eq:mass_cons2}),  
we can obtain the derivative $dp/dr$ described by the derivatives 
$du^r/dr$ and $d\ell/dr$ 
as 
\begin{equation}
-\frac{d\ln p}{dr}=P_r+P_u\frac{du^r}{dr}+P_\ell\frac{d\ell}{dr}.  
\label{eq:dp_urell}
\end{equation}
Usually, the derivation for the boundary condition at the sonic point 
become lengthy and complex for the general relativistic accretion flows. 
Eq. (\ref{eq:dp_urell}) is the key equation 
in order to simply treat the boundary condition at the sonic point. 
In the case of the general equation of state where the pressure 
is a function of 
both the rest-mass density and the temperature, 
if we use the energy equation 
given by Eq. (\ref{eq:energy}), 
we can also have the derivative $dp/dr$ with the form 
as Eq. (\ref{eq:dp_urell}). 
The examples of $P_r$, $P_u$ and $P_\ell$ for the isothermal disk, 
the polytropic disk and 
the ADAF with the general relativistic equation of state are given 
in Sec. \ref{sec:NSs}. 
From Eqs. (\ref{eq:n_HeatInertia}), (\ref{eq:shear0}) 
and (\ref{eq:diss_func}), 
the term $n_{\rm HI}$ containing the effects of the heat inertia  
is calculated as 
\begin{equation}
n_{\rm HI}=n_{\rm HI}^r+n_{\rm HI}^u\frac{du^r}{dr}
+n_{\rm HI}^\ell\frac{d\ell}{dr}, 
\label{eq:nHI_urell}
\end{equation} 
where 
\begin{eqnarray}
n_{\rm HI}^r\equiv\frac{u^r S}{\rho_0\eta}\sigma_r,~~
n_{\rm HI}^u\equiv\frac{u^r S}{\rho_0\eta}\sigma_u,~~
n_{\rm HI}^\ell\equiv\frac{u^r S}{\rho_0\eta}\sigma_\ell.  
\end{eqnarray}
Here, $S$ is calculated from Eqs. (\ref{eq:angmom2}) 
and (\ref{eq:tr_phi}) as 
$S=-\rho_0 u^r(\eta\ell-j)/F$. 
If the heat inertia effects by the flux is required, we change 
$n_{\rm HI}^r$ and $n_{\rm HI}^u$ according to Eq. (\ref{eq:nHI_b}). 
On the other hand, $\mathcal{D}_v$, $\mathcal{N}_v^r$ 
and $\mathcal{N}_v^u$ are defined  
in the next section in order to have $d\ell/dr$ with the form described as 
%
\begin{equation}
\mathcal{D}_v\frac{d\ell}{dr}
=\mathcal{N}_v^r+\mathcal{N}_v^u\frac{du^r}{dr}. \label{eq:dell_DvNv}
\end{equation}
%

%
By substituting $dp/dr$ and $n_{\rm HI}$ described 
by Eq. (\ref{eq:dp_urell}) and 
(\ref{eq:nHI_urell}), respectively, into the radial momentum equation 
given by 
Eq. (\ref{eq:radmom}), 
the derivative $du^r/dr$ can be calculated as 
\begin{equation}
\frac{du^r}{dr}=\frac{\mathcal{N}_s}{\mathcal{D}_s}, \label{eq:durdr_general}
\end{equation}
where
\begin{eqnarray}
\mathcal{D}_s&=& u^r
-h^{rr}c_s^2
	\left(P_u+P_\ell \frac{\mathcal{N}_v^u}{\mathcal{D}_v}\right)
-\left(n_{\rm HI}^u+n_{\rm HI}^\ell 
\frac{\mathcal{N}_v^u}{\mathcal{D}_v} \right),
\\
\mathcal{N}_s&=& 
n_{\rm acc}
+h^{rr}c_s^2
	\left(P_r+P_\ell \frac{\mathcal{N}_v^r}{\mathcal{D}_v}\right)
+\left(n_{\rm HI}^r+n_{\rm HI}^\ell 
\frac{\mathcal{N}_v^r}{\mathcal{D}_v} \right).
\end{eqnarray}
Here, we use Eq. (\ref{eq:dell_DvNv}) to remove the derivative $d\ell/dr$. 
In order to pass the sonic point smoothly where $\mathcal{D}_s=0$, 
the condition $\mathcal{N}_s=0$ must be satisfied at the sonic point. 
So, the boundary conditions at the sonic point are 
\begin{equation}
\mathcal{D}_s=\mathcal{N}_s=0. 
\end{equation}
%

\subsection{Another Boundary Condition}
\label{sec:VP}
\subsubsection{For Type A Causal Viscosity: Boundary Condition at Horizon}
For the type 1 causal viscosity prescription, 
we put the boundary condition at 
the horizon in order to vanish 
the angular momentum transportation at the horizon. 
At $r=r_+$, from the condition $\nu=0$, 
$\eta\ell=j$ is required and is used as 
the boundary condition. 
In this case, the parameters given 
at the sonic point determine the transonic solution. 
The differential equation for the angular momentum, 
$\ell$, is calculated from 
the equation for the angular momentum conservation and 
the Navier-Stokes prescription 
of the viscosity as $t_{\mu\nu}=-2\rho_0\eta\nu\sigma_{\mu\nu}$ as  
\begin{equation}
\frac{d\ell}{dr}
	=\frac{1}{h^{rr}}
		\left(
			2\sigma^r_{~\phi}-u^r_{~;\phi}
			-g^{tr}u_{\phi;t}
			-g^{r\phi}u_{\phi;\phi}
			+n_{\rm acc}\ell+g^{rr} \Gamma_{\phi r}^\mu u_\mu
			+\frac{2}{3}\Theta_r u^r \ell
			-\frac{u^r\ell}{3}\frac{du^r}{dr}
		\right), 
\end{equation}
where 
\begin{eqnarray}
u^r_{~;\phi}&=&
	\frac{1}{2}(g^{tr}g_{t\phi,r}+g^{rr}g_{r\phi,r}
	+g^{r\phi}g_{\phi\phi,r})u^r
	-\frac{1}{2}g^{rr}(g_{t\phi,r}+g_{\phi\phi,r}\Omega)u^t, \\
u_{\phi;t}&=&\frac{1}{2}g_{t\phi,r}u^r,\\
u_{\phi\phi}&=&\frac{1}{2}g_{\phi\phi,r}u^r,\\
\Gamma_{\phi r}^\mu u_\mu &=& \frac{1}{2}(g_{t\phi,r}+g_{\phi\phi,r}\Omega)u^t,  
\end{eqnarray}
and $\Theta_r$ is given in Appendix. 
Here, the shear rate $\sigma^r_{~\phi}$ is calculated as 
\begin{equation}
\sigma^r_{~\phi}=\frac{u^r(\eta\ell-j)}{2\eta\nu}, 
\end{equation}
when the value of the kinematic viscosity coefficient is not zero, i.e., 
$\nu\neq 0$. 
When $\nu=0$, this shear rate is zero, i.e. $\sigma^r_{~\phi}=0$.  

\subsubsection{For Type B Causal Viscosity: Boundary Conditions at 
Viscous Point}
The boundary conditions at the viscous point are calculated from the equation
for $d\ell/dr$. 
In the same way as Eq. (\ref{eq:dp_urell}), 
the derivatives of $\eta$, $H_\theta$ and $F$ with respect to $r$ are 
described by the combinations of $du^r/dr$ and $d\ell/dr$ as 
\begin{eqnarray}
\frac{d\ln\eta}{dr}&=&\eta_r+\eta_u\frac{du^r}{dr}
	+\eta_\ell\frac{d\ell}{dr},
\label{eq:deta_urell}\\
\frac{d\ln H_\theta}{dr}&=&H_r+H_u\frac{du^r}{dr}+H_\ell\frac{d\ell}{dr},
\label{eq:dH_urell}\\
\frac{d\ln F}{dr}&=&F_r+F_u\frac{du^r}{dr}+F_\ell\frac{d\ell}{dr}. 
\label{eq:F_urell} 
\end{eqnarray}
By substituting Eqs. (\ref{eq:S0}), (\ref{eq:shear0}), 
(\ref{eq:deta_urell}), (\ref{eq:dH_urell}) and (\ref{eq:F_urell}) 
into Eq. (\ref{eq:S}) and using 
the relation $S=-\rho_0 u^r(\eta\ell-j-Q_\ell)/F$, 
the equation for $d\ell/dr$ containing no derivatives except $du^r/dr$ 
can be calculated as, 
%
\begin{equation}
\frac{d\ell}{dr}=\frac{\mathcal{N}_v}{\mathcal{D}_v},
\label{eq:delldr_general}
\end{equation}
%
where $\mathcal{N}_v=\mathcal{N}_{v}^r+\mathcal{N}_{v}^u (du^r/dr)$ and 
%
\begin{eqnarray}
\mathcal{D}_v&=&
	1-\frac{(u^r)^2}{2 \sigma_\ell F c_v^2}
		\left[
			1+\ell\eta_\ell-\frac{\tilde{Q}_\ell}{\eta}
			+\frac{SF}{\rho_0 \eta u^r}G_\ell
		\right], \\
\mathcal{N}_{v}^r&=&
	-\frac{\sigma_r}{\sigma_\ell}
	+\frac{(u^r)^2}{2\sigma_\ell F c_v^2}
		\left[\ell\eta_r-\frac{\tilde{Q}_r}{\eta}
		+\frac{SF}{\rho_0 \eta u^r}\left(G_r-\frac{1}{u^r\tau_v}\right)
		\right],
\\
\mathcal{N}_{v}^u&=&
	-\frac{\sigma_u}{\sigma_\ell}
	+\frac{(u^r)^2}{2\sigma_\ell F c_v^2}
		\left[\ell\eta_u-\frac{\tilde{Q}_u}{\eta}
			+\frac{SF}{\rho_0\eta u^r}G_u
		\right]. 
\end{eqnarray}
%
Here, 
$G_r$, $G_u$ and $G_\ell$ are defined as
\begin{equation}
G_r\equiv 2/r+F_r+H_r,~
G_u\equiv F_u+H_u,~
G_\ell\equiv F_\ell+H_\ell,    
\end{equation} 
in order to have the relation 
\begin{equation}
\frac{2}{r}+\frac{d\ln F}{dr}+\frac{d\ln H_\theta}{dr}
=G_r+G_u\frac{du^r}{dr}+G_\ell\frac{d\ell}{dr}, 
\end{equation}
in the denominator of Eq. (\ref{eq:S}), 
and $\tilde{Q}_r$, $\tilde{Q}_u$ and $\tilde{Q}_\ell$ are determined in 
order to have the relation 
\begin{equation}
\frac{dQ_\ell}{dr}=\tilde{Q}_r+\tilde{Q}_u\frac{du^r}{dr}
+\tilde{Q}_\ell\frac{d\ell}{dr}.   
\end{equation} 
In this study, we neglect the angular momentum loss by the radiation, i.e. 
we set 
$\tilde{Q}_r=\tilde{Q}_u=\tilde{Q}_\ell=0$. 
In order to pass the viscous point smoothly where $\mathcal{D}_v=0$, 
the condition $\mathcal{N}_v=0$ must be satisfied at the viscous point. 
So, the boundary conditions at the viscous point are 
\begin{equation}
\mathcal{D}_v=\mathcal{N}_v=0. 
\end{equation}
%

\section{Coupled Differential Equations to be solved}
\label{sec:coupled_eq}
For the general equation of state, the transonic solutions are obtained by 
numerically solving the coupled differential equations 
for the dynamic variables, 
e.g. $u^r$, $\ell$, and the thermodynamic variables, e.g. $T$. 
In the case of the special thermodynamic relations, such 
as the isothermal flows and 
the polytropic flows, the thermodynamic variables can be calculated from 
the dynamical variables. 
In these cases, we only solve the coupled differential equation for 
the dynamic variables. 
In this study, we treat the radial component of the four velocity, 
$u^r$, and the 
angular momentum, $\ell$, as the basic dynamical variables to be solved. 
That is, for the case of the special thermodynamic relations 
where the thermodynamic variables 
are calculated from the dynamic variables, we solve 
the coupled differential equations 
for $u^r$ and $\ell$ described as 
%
\begin{eqnarray}
\frac{du^r}{dr}&=&\frac{\mathcal{N}_s}{\mathcal{D}_s}, \label{eq:durdr}\\
\frac{d\ell}{dr}&=&\frac{\mathcal{N}_v}{\mathcal{D}_v}~~ 
	\left(=\frac{\mathcal{N}_v^r}{\mathcal{D}_v}
	+\frac{du^r}{dr}\frac{\mathcal{N}_v^u}{\mathcal{D}_v}\right). 
\label{eq:delldr}
\end{eqnarray}
%
In the following sections where the transonic solutions for 
the isothermal flows and the polytropic flows are calculated, 
we solve these two differential equations. 
On the other hand, for the general equation of state, 
the differential equation for the thermodynamic variables 
is usually solved in addition to 
Eqs. (\ref{eq:durdr}) and (\ref{eq:delldr}). 
The differential equations for the thermodynamic variables 
are derived by using the energy 
equation given by Eq. (\ref{eq:energy}). 
In this study, we treat the temperature $T$ as the basic 
thermodynamic variable whose differential 
equation is numerically solved. 
The other thermodynamic variables, such as the rest-mass density 
$\rho_0$, the sound velocity $c_s$, 
are calculated from $u^r$, $\ell$ and $T$ by using 
the mass conservation equation given by Eq. 
(\ref{eq:mass_cons2}) and the equation of state. 
Here, we derive the general form of the differential equation 
for $T$ by using the energy equation.  
By differentiating the mass conservation 
given Eq. (\ref{eq:mass_cons2}), the disk thickness 
$H_\theta=c_s r/\ell_*$ and the sound velocity 
$c_s=[p/(\eta\rho_0)]^{1/2}$ with respect 
to $r$, we obtain 
\begin{eqnarray}
&&
\frac{2}{r}
	+\frac{d\ln\rho_0}{dr}
	+\frac{d\ln|u^r|}{dr}
	+\frac{d\ln H_\theta}{dr}
	=0,\label{eq:thermo1}\\
&&
\frac{d\ln H_\theta}{dr}
	=\frac{d\ln c_s}{dr}
	+\frac{1}{r}
	-\frac{1}{\ell_*}\frac{d\ell_*}{dr}
	,\label{eq:thermo2}\\
&&
2\frac{d\ln c_s}{dr}
	=\frac{d\ln p}{dr}
	-\frac{d\ln \rho_0}{dr}
	+\frac{d\ln \eta}{dr}. \label{eq:thermo3}
\end{eqnarray}
Here $d\ell_*/dr$ can be written by the linear combination 
of $du^r/dr$ and $d\ell/dr$ as 
\begin{equation}
\frac{d\ell_*}{dr}=\ell_*^r+\ell_*^u
\frac{du^r}{dr}+\ell_*^\ell\frac{d\ell}{dr}, 
\end{equation}
where the coefficients $\ell_*^r$, $\ell_*^u$ and $\ell_*^\ell$ 
can be calculated analytically or 
numerically. 
We newly define 
%
\begin{equation}
P_\rho \equiv \left(\frac{\partial \ln p}{\partial \ln \rho_0}\right)_T,~~
	P_T \equiv \left(\frac{\partial \ln p}{\partial \ln T}\right)_{\rho_0},~~
U_\rho \equiv \left(\frac{\partial \ln u}{\partial \ln \rho_0}\right)_T,~~
U_T \equiv 
	\left(\frac{\partial \ln u}{\partial \ln T}\right)_{\rho_0},~~
\eta_\rho \equiv 
	\left(\frac{\partial \ln \eta}{\partial \ln \rho_0}\right)_T,~~
\eta_T \equiv 
	\left(\frac{\partial \ln \eta}{\partial \ln T}\right)_{\rho_0}. 
\end{equation}
Here, $\eta_\rho$ and $\eta_T$ are related to 
$P_\rho$, $P_T$, $U_\rho$ and $U_T$ as 
$\eta_\rho=(uU_\rho+pP_\rho-u-p)/(\eta\rho_0)$ 
and $\eta_T=(uU_T+pP_T)/(\eta\rho_0)$. 
From Eqs. (\ref{eq:dp_rho0T}), (\ref{eq:du_rho0T}), 
(\ref{eq:thermo1}), (\ref{eq:thermo2}) and (\ref{eq:thermo3}), 
we obtain the differential equation for $T$ and $\rho_0$ as 
\begin{eqnarray}
\frac{d\ln T}{dr}&=&T_r+T_u\frac{du^r}{dr}+T_\ell\frac{d\ell}{dr},
\label{eq:dTdr}\\
\frac{d\ln \rho_0}{dr}&=&\rho_r
+\rho_u\frac{du^r}{dr}+\rho_\ell\frac{d\ell}{dr}. 
\end{eqnarray}
Here $T_k$ and $\rho_k$ ($k=r$, $u$ and $\ell$) are calculated as 
\begin{eqnarray}
T_k&=&\mathcal{C}^T_1 \mathcal{X}_k +\mathcal{C}^T_2 q^\pm_k/u^r, 
\label{eq:T_k}\\
\rho_k&=&\mathcal{C}^\rho_1 \mathcal{X}_k +\mathcal{C}^\rho_2 q^\pm_k/u^r,
\end{eqnarray}
where $\mathcal{X}_k$ ($k=r$, $u$ and $\ell$) are given as  
\begin{equation}
\mathcal{X}_r=\frac{3}{r}-\frac{\ell_*^r}{\ell_*},~
\mathcal{X}_u=\frac{1}{u^r}-\frac{\ell_*^u}{\ell_*},~
\mathcal{X}_\ell=-\frac{\ell_*^\ell}{\ell_*}, 
\end{equation}
and $\mathcal{X}_T^1$, $\mathcal{X}_T^2$, $\mathcal{X}_\rho^1$ 
and $\mathcal{X}_\rho^2$ 
are given as
\begin{eqnarray}
\mathcal{C}^T_1=\frac{2}{\mathcal{X}_D}\left(U_\rho-\frac{u+p}{u}\right),~~
\mathcal{C}^T_2=\frac{1}{u\mathcal{X}_D}(P_\rho-\eta_\rho+1),~~
\mathcal{C}^\rho_1=-\frac{2U_T}{\mathcal{X}_D},~~
\mathcal{C}^\rho_2=-\frac{(P_T-\eta_T)}{u\mathcal{X}_D},~~
\end{eqnarray}
Here, $\mathcal{X}_D$ is defined as 
\begin{eqnarray}
\mathcal{X}_D&\equiv& (P_\rho-\eta_\rho+1)U_T-(P_T-\eta_T)
\left(U_\rho-\frac{u+p}{u}\right).  
\end{eqnarray}
On the other hand, 
$q^\pm_k$ ($k=r$, $u$ and $\ell$) are defined to satisfy the relation 
\begin{equation}
q_{\rm vis}^+ -q_{\rm rad}^-
=q^\pm_r+q^\pm_u \frac{du^r}{dr}+q^\pm_\ell \frac{d\ell}{dr}. 
\end{equation}
By using the coefficients $T_k$ and $\rho_k$ ($k=r$, $u$ and $\ell$) 
calculated above, the coefficients $P_k$, $\eta_k$ and $H_k$ 
($k=r$, $u$ and $\ell$) 
are calculated as 
\begin{eqnarray}
P_k&=&-P_T T_k-P_\rho \rho_k, \label{eq:P_k}\\
\eta_k&=&\eta_T T_k+\eta_\rho \rho_k. \label{eq:eta_k}
\end{eqnarray}
From Eq. (\ref{eq:thermo1}), we obtain 
%
\begin{equation}  
H_r=-\frac{2}{r}-\frac{\rho_r}{\rho_0},~
H_u=-\frac{1}{u^r}-\frac{\rho_u}{\rho_0},~
H_\ell=-\frac{\rho_\ell}{\rho_0}.\label{eq:H_k} 
\end{equation}
For the general equation of state, the transonic solutions are obtained by 
solving the differential equations of $u^r$, $\ell$ and 
$T$ given by Eqs. (\ref{eq:durdr}), (\ref{eq:delldr}) and (\ref{eq:dTdr}). 
In the later sections, for the ADAF with the general relativistic 
equation of state and the 
supercritical accretion disk, we solve the differential 
equations for $u^r$, $\ell$ and $T$. 
Since the rest-mass density $\rho_0$ is calculated 
from the mass conservation equation 
given by Eq. (\ref{eq:mass_cons2}), we do not solve 
the differential equation for $\rho_0$. 
%

\section{Calculation Method}
\label{sec:calc_method}
By using the formalism developed until the last sections, 
we solve the coupled differential equations to obtain 
the transonic solutions. 
The calculation method for the transonic solutions are not unique, and 
actually, past studies use several method. 
Here, we show one of the calculation methods to obtain 
the transonic solutions.  
%
\begin{enumerate}
\item First, we tentatively choose some value of $T_s$ (or $a_{s,s}$) 
for given values of $r_s$ and $j$, and  
calculate $u^r_s$, $\ell_s$, $(du^r/dr)_s$, $(d\ell/dr)_s$ and $(dT/dr)_s$.  
Here, the differential values at the sonic point 
are calculated by using the L'Hopital's rule. 
\item Next, we solve the solutions in the range $r_s<r<r_v$. 
In order to do this, we solve the coupled differential equations 
for $u^r$, $\ell$ and $T$  
from the sonic point to the viscous point by using, e.g., 
the Runge-Kutta algorithm. 
Usually, for the initially selected value of $T_s$ (or $a_{s,s}$), 
the calculated solution does not pass the 
viscous point where two boundary conditions 
$\mathcal{D}_v=\mathcal{N}_v=0$ are satisfied. 
In such case, we return to step 1 and again choose the different values of 
$T_s$ (or $a_{s,s}$) for given values of $r_s$ and $j$. 
After repeating these procedures, 
we can determine the value $T_s$ (or $a_{s,s}$) which gives the solution 
satisfying the boundary conditions $\mathcal{D}_s=\mathcal{N}_s=0$ 
at $r=r_s$ and 
$\mathcal{D}_v=\mathcal{N}_v=0$ at $r=r_v$. 
\item After solving the solutions in $r_s<r<r_v$, 
we solve the coupled differential equations in the range $r_v<r$ 
by using the values of $T_s$ (or $a_{s,s}$) for given values of 
$r_s$ and $j$ by using, e.g., the Runge-Kutta algorithm. 
\item Finally, we solve the coupled differential equations 
in the range $r<r_s$ by using the values of $T_s$ (or $a_{s,s}$) 
for given values of 
$r_s$ and $j$ by using, e.g., the Runge-Kutta algorithm. 
If the solutions are connected with the horizon as usual solutions, 
we can solve the solutions inside the horizon. 
On the other hand, if the solutions are not connected with 
the horizon as the alpha-type solutions, 
the numerical integrations are stopped before the horizon 
because there is no stationary solutions 
for such parameters of $r_s$ and $j$.  
%
\end{enumerate}
%
%
The third step and the fourth step can be interchanged. 
By this procedure, the transonic solutions are obtained 
for given values of $r_s$ and $j$ 
without the boundary conditions for the inner regions ($r<r_s$) 
of the outer regions ($r_v<r$). 
By using these procedures, we can basically cover 
all the possible values of $r_s$ and $j$. 
That is, by these methods, in principle, 
all the possible stationary transonic solutions can be calculated 
because we can use all the possible sonic point, and 
the transonic solution is 
calculated from the sonic point. 
%

%
\section{Applications and Sample Solutions}
\label{sec:NSs}
In this section, we give the numerical solutions for 
the ideal isothermal accretion flow (\S \ref{sec:IdIs}), 
the polytropic disks (\S \ref{sec:PD}), 
the ADAF with relativistic equation of state (\S \ref{sec:ADAF}) and 
the adiabatic accretion flow (\S \ref{sec:AD}) and 
the formulation for the supercritical accretion flow (\S \ref{sec:SCD}). 
Based on the thermodynamic relations,  
we first calculate the coefficients 
$P_k$, $\eta_k$ and $H_k$ ($k=r$, $u$ and $\ell$)  
which are used in the calculations of $du^r/dr$ and $d\ell/dr$. 
In addition, we calculate $T_k$ ($k=r$, $u$ and $\ell$) if required. 
For the ideal flows, we only solve the differential equation of $u^r$. 
For the viscous polytropic flows, we solve the coupled differential 
equations of $u^r$ and $\ell$. 
For ADAFs with the relativistic equation of state, 
in addition to the differential equation of $u^r$ and $\ell$, 
we also solve the differential equation of $T$ simultaneously, 
which is derived by using the energy equation. 
In this section, for the viscous solutions, 
the kinematic viscosity $\nu$ is assumed to described 
by the alpha viscosity $\alpha_v$
as $\nu=\alpha_v c_s^2/\Omega$. 
%

\subsection{Application 1 : Ideal Isothermal Accretion Flow}
\label{sec:IdIs}
\begin{figure}
\begin{center}
\includegraphics[width=170mm]{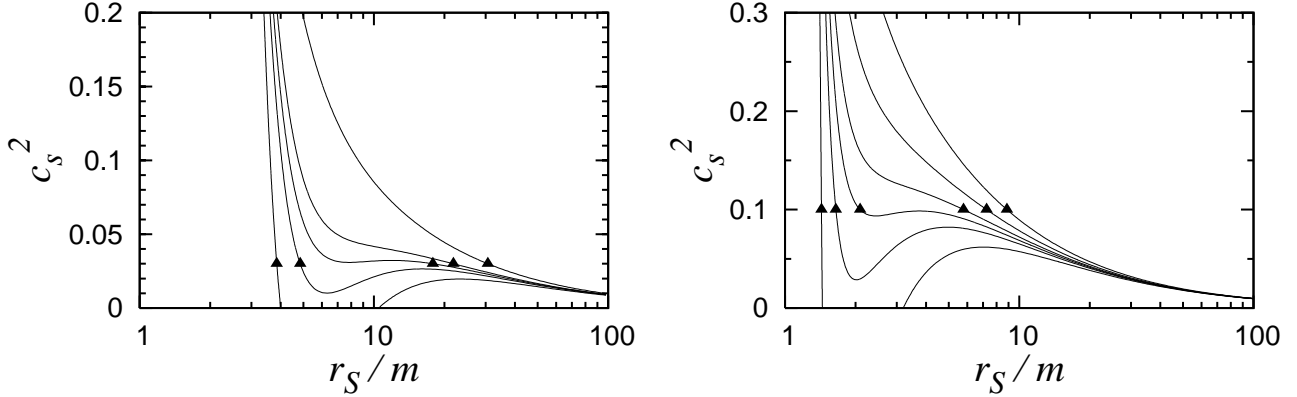}
\end{center}
\caption{
Parameter spaces $r_S$-$c_s^2$ showing 
lines of constant critical values of 
$\lambda(\equiv -u_\phi/u_t)$. 
The critical sound speed $c_s^2$ is plotted 
for non-rotating ($a/m=0$:{\it left panel}) 
and rotating ($0.95$: {\it right panel}) black holes.
For non-rotating black holes, 
contours correspond to $\lambda=2.0$, 3.2, 3.4, 3.6 and 4.0 
(from right to left). 
For rotating black holes, contours correspond 
to $\lambda=1.5$, 1.9, 2.1, 2.2, 2.3 and 2.5
(from right to left). 
Transonic solutions with the critical values plotted 
by the filled triangles are 
calculated in Fig. \ref{fig:isid_a00} for non-rotating black holes and in 
Fig. \ref{fig:isid_a095} for rotating black holes. 
}
\label{fig:rscs2}
\end{figure}

\begin{figure*}
\includegraphics[width=170mm]{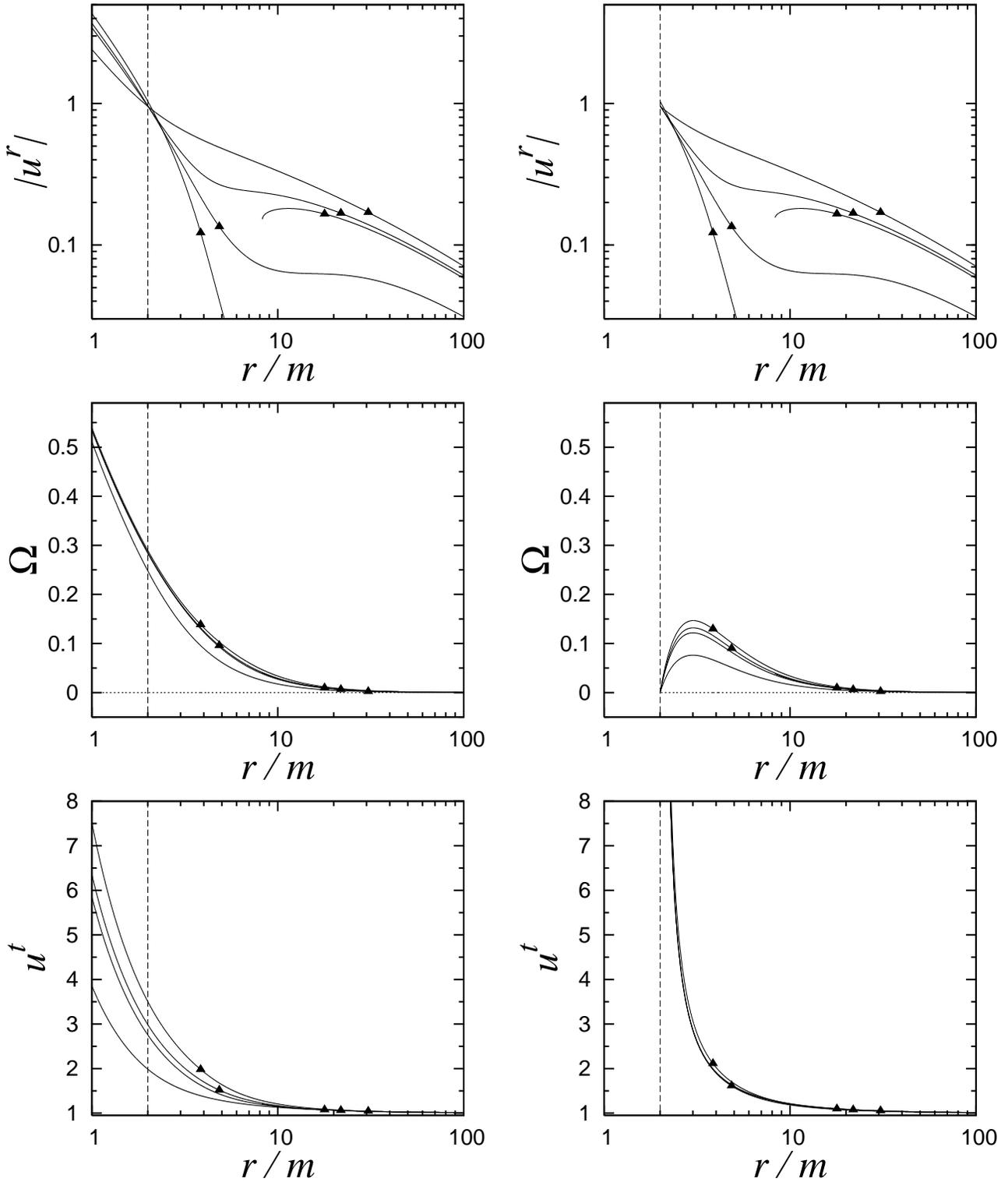}
\caption{
Sample transonic solutions for the ideal isothermal flows when $a/m=0$. 
We plot transonic solutions with the critical values 
$\lambda=2.0$, 3.2, 3.4, 3.6 and 4.0 (from right to left) 
with the sound speed $c_s^2=0.03$. 
The transonic solutions calculated by using the 
Kerr-Schild coordinate ({\it left column}) 
or the Boyer-Lindquist coordinate ({\it right column}) are plotted.   
The radius of the horizon is denoted by the dashed line, and 
the angular velocity of the 
frame dragging is plotted by the dotted lines in the panel 
showing $\Omega$.  
}
\label{fig:isid_a00}
\end{figure*}

\begin{figure*}
\includegraphics[width=170mm]{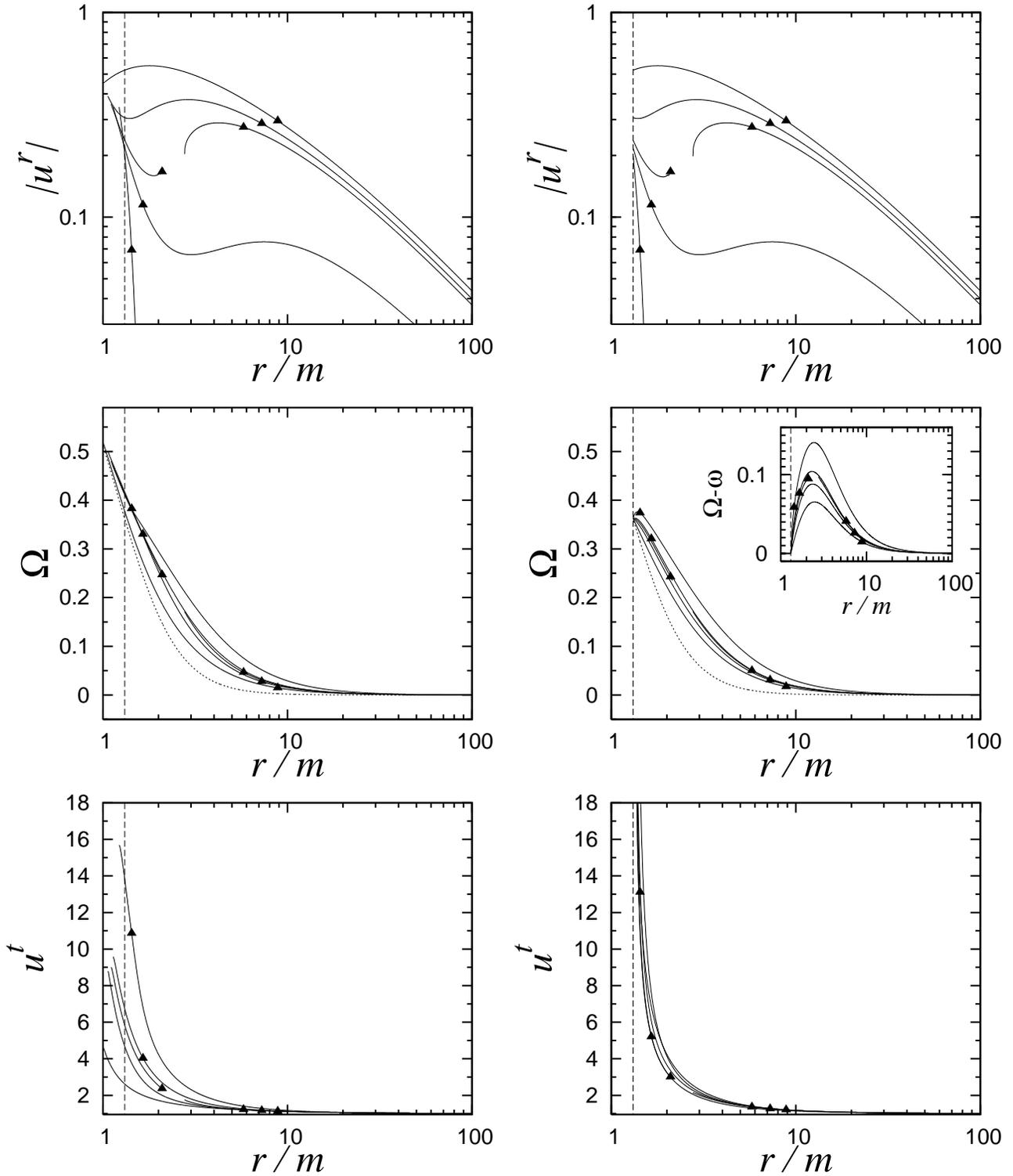}
\caption{
Sample transonic solutions for the isothermal ideal flows when $a/m=0.95$. 
We plot transonic solutions with the critical values 
$\lambda=1.5$, 1.9, 2.1, 2.2, 2.3 and 2.5 (from right to left) 
with the sound speed $c_s^2=0.1$. 
The transonic solutions calculated by using 
the Kerr-Schild coordinate ({\it left column}) 
or the Boyer-Lindquist coordinate ({\it right column}) are plotted. 
The radius of the horizon is denoted by the dashed line, 
and the angular velocity of the 
frame dragging is plotted by the dotted lines in the panel showing $\Omega$. 
We also show $\Omega-\omega$ in the inserted box 
in the panel showing $\Omega$ in the right column. 
}
\label{fig:isid_a095}
\end{figure*}

By using the formalism developed until the last sections, 
we first shows the numerical solutions of the horizon-penetrating solutions 
for the ideal isothermal accretion flow which is one of 
the simplest transonic accretion flow. 
Here, we only solve the differential equation for $u^r$ 
by assuming constant specific angular 
momentum $\ell$ and sound speed $c_s$, and $\eta=1$ is also assumed. 
We use the coefficients $P_r$, $P_u$ and $P_\ell$ described as 
\begin{equation}
P_r=\frac{3}{r},~~P_u=\frac{1}{u^r},~~P_\ell=0.  \label{eq:Pk_IdIs}
\end{equation}
By substituting Eqs. (\ref{eq:Pk_IdIs}) into Eqs. (\ref{eq:durdr_general}), 
the differential equation for $u^r$ for the ideal isothermal flows 
are obtained. 
We numerically solve this differential equation for $u^r$ and 
obtain the transonic 
solutions. 
When calculating the numerical solutions, the rest-mass density 
$\rho_0$ is determined 
from $u^r$, $\ell$ and $c_s$ by using the mass conservation 
equation given by 
Eq. (\ref{eq:mass_cons2}). 
For the ideal isothermal flows which are solved in this section, 
since we assume constant $\ell$, there is no viscous point 
in the global solution of the transonic 
accretion flow. 
In Fig. \ref{fig:rscs2}, in the parameter spaces $r_s$-$c_s^2$ we plot 
lines of constant critical values of $\lambda(\equiv -u_\phi/u_t)$. 
The critical sound speed $c_s^2$ is plotted for non-rotating 
($a/m=0$:{\it left panel}) 
and rotating ($0.95$: {\it right panel}) black holes.
We calculate transonic solutions with the critical values plotted 
by the filled triangles 
in Fig. \ref{fig:rscs2}. 
The resultant transonic solutions for non-rotating black holes are plotted 
in Fig. \ref{fig:isid_a00}. 
We also give the transonic solutions for rotating black holes 
in Fig. \ref{fig:isid_a095}. 
For both Fig. \ref{fig:isid_a00} and Fig. \ref{fig:isid_a095}, 
the transonic solutions calculated by using the Kerr-Schild 
coordinate ({\it left column}) 
or the Boyer-Lindquist coordinate ({\it right column}) are plotted. 
The angular velocity $\Omega$ and $u^t$ directly reflect the effects of the 
coordinate singularity when we use the Boyer-Lindquist coordinate.  
That is, for the solutions calculated in the 
Boyer-Lindquist coordinate, the angular velocity, 
$\Omega$, is equal to the angular velocity of the frame dragging, $\omega$, 
at the horizon, and $u^t$ is diverged at the horizon. 
These features are clearly seen in both Fig. \ref{fig:isid_a00} 
and Fig. \ref{fig:isid_a095}. 
We also show $\Omega-\omega$ in the inserted box 
in the panel showing $\Omega$ in the right 
column of Fig. \ref{fig:isid_a095}. 
The feature that the angular velocity of the accretion flow written 
by the Boyer-Lindquist 
coordinate is equal to the angular velocity of the black hole 
$\omega=a/2mr$ at the horizon  
is pointed out by \cite{k04} who also found that $u^t$ remains 
finite at the event horizon 
and $\Omega$ differs from the angular velocity of the black hole 
in Kerr-Schild coordinate. 
The coordinate singularity in Boyer-Lindquist coordinate 
is also relevant to the feature 
that the world lines of Boyer-Lindquist LNRF become null 
on the event horizon and thus 
cannot correspond to any physical observer. 
For the outside region of the horizon,  
the lines for $u^r$ are same for both calculations using 
Kerr-Schild coordinate and the 
Boyer-Lindquist coordinate. 
We plot two types of transonic solutions which have the sonic radius 
in the inside region 
or the outside region. 
These two types of solutions correspond to the solutions named type I 
and type II in 
Peitz \& Appl (1997). 
Similar solution patterns are also obtained by \cite{f87}. 
For the accretion flows calculated by using the Boyer-Lindquist coordinate, 
we also plot the results for the flows which are firstly calculated 
based on the Boyer-Lindquist 
coordinate and then transformed to the flows written by 
the Kerr-Schild coordinate 
by the transformations of four velocity given by Eqs. (\ref{eq:uBLKS}) 
and (\ref{eq:uBLKS2}). 
These results are plotted by the short dashed lines in the right panels 
for $\Omega$ and $u^t$ 
of Figs. \ref{fig:isid_a00} and Fig. \ref{fig:isid_a095}. 
These solutions outside the event horizon are same as 
those calculated based on the Kerr-Schild 
coordinate presented in the left panels of Figs. \ref{fig:isid_a00} 
and Fig. \ref{fig:isid_a095}. 

Outside the horizon, from the results for 
the accretion flow calculated by using the 
Boyer-Lindquist coordinate shown in the right panels of 
Figs. \ref{fig:isid_a00} and \ref{fig:isid_a095}, 
we can obtain the solutions given in the left panels 
for the Kerr-Schild coordinate by 
using the transformation given by Eqs. (\ref{eq:uBLKS}) 
and (\ref{eq:uBLKS2}). 
While for the ideal accretion flows 
the accretion flows calculated by these two procedures 
have same results, for the viscous flows the solutions 
calculated by these two procedures do not have 
the exactly same results. 
See the discussion in the last section. 
%

%
%

\begin{figure*}
\includegraphics[width=170mm]{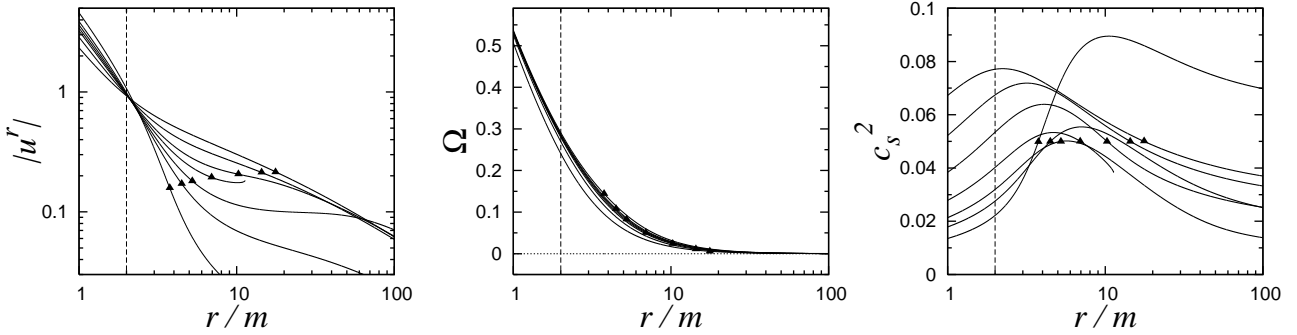}
\caption{
Sample transonic solutions for the polytropic ideal flows for $a/m=0$. 
We plot transonic solutions with the critical values 
$\lambda=2.0$, 2.6, 3.0, 3.2, 3.4, 3.6 and 4.0 (from right to left) 
with the critical sound speed $c_s^2=0.05$. 
The radius of the horizon is denoted by the dashed line, and the sonic points are 
plotted by the filled triangles.  
}
\label{fig:PDideal_a00}
\end{figure*}
\begin{figure*}
\includegraphics[width=170mm]{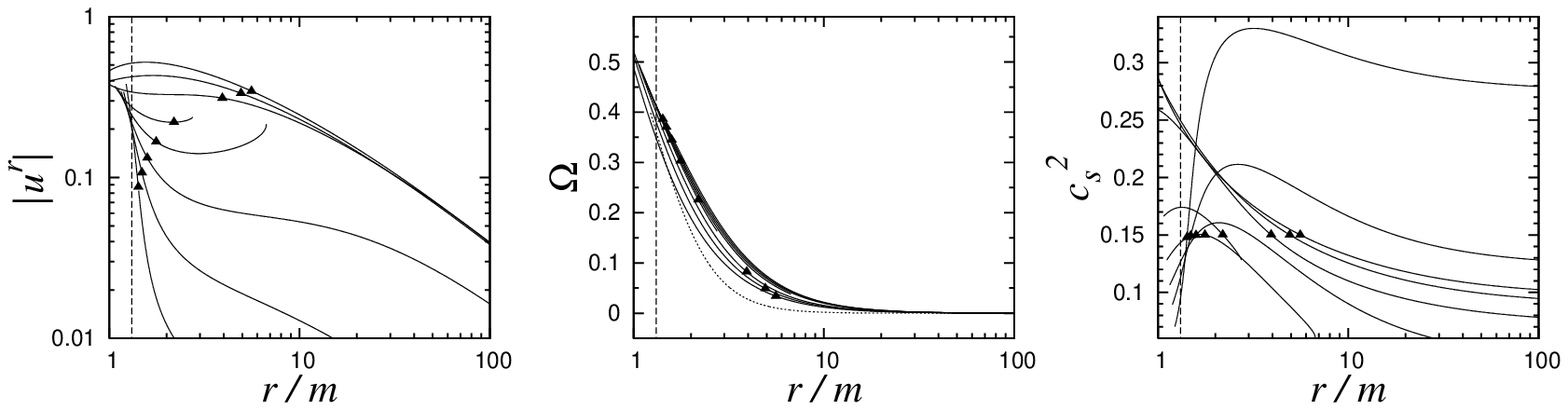}
\caption{
Sample transonic solutions for the polytropic ideal flows for $a/m=0.95$. 
We plot transonic solutions with the critical values 
$\lambda=1.5$, 1.7, 1.9, 2.1, 2.2, 2.3, 2.4 and 2.5 (from right to left) 
with the critical sound speed $c_s^2=0.15$. 
}
\label{fig:PDideal_a095}
\end{figure*}
%
\subsection{Application 2 : Polytropic Accretion Flow}
\label{sec:PD}
Here, we show the transonic accretion flows with the polytropic equation flows. 
%
%
Although this equation determine the general energy equation, 
in this paper, we only consider the accretion flows with the polytropic equation 
of state as 
%
\begin{equation}
p=K\rho_0^\Gamma, \label{eq:p_poly}
\end{equation}
%
where $K$ is constant and $\Gamma$ is the adiabatic index (or the ratio of specific heat). 
The adiabatic index $\Gamma$ is related to the polytropic index $N$ as 
$\Gamma=1+1/N$. 
The internal energy $u$ is given by $u=p/(\Gamma-1)$. 
When solving the transonic solutions for polytropic accretion flows, 
we do not use the energy equation in Eq. (\ref{eq:energy}).  
All the thermodynamic variables are 
expressed by the rest-mass density $\rho_0$ for given adiabatic index $\Gamma$ and 
the constant $K$. 
The relativistic enthalpy $\eta~[=1+(u+p)/\rho_0]$ 
and the sound speed $c_s~[=p/(\rho_0\eta)]$ are calculated as 
%
\begin{eqnarray}
\eta&=&1+\left(\frac{\Gamma}{\Gamma-1}\right)\frac{p}{\rho_0}, \label{eq:eta_poly}\\
c_s^2&=&\frac{p}{\rho_0+[\Gamma/(\Gamma-1)]p}. \label{eq:as_poly}
\end{eqnarray}
%
From Eqs. (\ref{eq:p_poly}), (\ref{eq:eta_poly}) and (\ref{eq:as_poly}), 
$dp/dr$, $d\eta/dr$ and $dc_s/dr$ and can be calculated from $d\rho_0/dr$ as
%
\begin{eqnarray}
\frac{dp}{dr}&=&\frac{\Gamma p}{\rho_0}\frac{d\rho_0}{dr}, \label{eq:dpdr_poly}\\
\frac{d\eta}{dr}&=&\frac{\Gamma p}{\rho_0^2}\frac{d\rho_0}{dr},\label{eq:detadr_poly}\\ 
\frac{dc_s}{dr}&=&\frac{(\Gamma-1)c_s}{2\eta\rho_0}\frac{d\rho_0}{dr}. \label{eq:dasdr_poly}
\end{eqnarray}
%
From the disk thickness given by Eq. (\ref{eq:Ht2KS}) and the mass conservation given by 
Eq. (\ref{eq:mass_cons2}), we obtain Eqs. (\ref{eq:thermo1}) and (\ref{eq:thermo2}).  
By eliminating $dH_\theta/dr$ from Eqs. (\ref{eq:thermo1}) and (\ref{eq:thermo2}) 
and substituting Eq. (\ref{eq:dasdr_poly}), 
the derivative $d\rho_0/dr$ is calculated as, 
\begin{equation}
\frac{d\rho_0}{dr}=
\frac{-2\eta\rho_0}{2\eta+\Gamma-1}
\Big[
\frac{4}{r}-\frac{\ell_*^r}{\ell_*}
+\left(\frac{1}{u^r}-\frac{\ell_*^u}{\ell_*}\right)\frac{du^r}{dr}
-\frac{\ell_*^\ell}{\ell_*}\frac{d\ell}{dr}
\Big]. \label{eq:drhodr_poly}
\end{equation}
By substituting Eq. (\ref{eq:drhodr_poly}) into Eq. (\ref{eq:dpdr_poly}), 
$P_r$, $P_u$ and $P_\ell$ are calculated as, 
\begin{equation}
P_r=\mathcal{X}_p \Big(\frac{4}{r}-\frac{\ell_*^r}{\ell_*}\Big),~
P_u=\mathcal{X}_p\Big(\frac{1}{u^r}-\frac{\ell_*^u}{\ell_*}\Big),~
P_\ell=-\mathcal{X}_p\frac{\ell_*^\ell}{\ell_*}. 
	\label{eq:Pruell_poly}
\end{equation}
Here, we define $\mathcal{X}_p\equiv 2\eta\Gamma/(2\eta+\Gamma-1)$. 
In the same way, $\eta_r$, $\eta_u$ and $\eta_\ell$ are calculated as,
\begin{equation}
\eta_r=-c_s^2 P_r,~~\eta_u=-c_s^2 P_u,~~\eta_\ell=-c_s^2 P_\ell,  
\label{eq:eta_ruell_poly}
\end{equation}
and 
$H_r$, $H_u$ and $H_\ell$ are calculated as, 
%
\begin{eqnarray}
H_r=-\mathcal{X}_p
	\Big(\frac{\Gamma-1}{\eta r}-\frac{1}{r}+\frac{\ell_*^r}{\ell_*}\Big),~~~
H_u=-\mathcal{X}_p
	\Big(\frac{\Gamma-1}{2\eta u^r}+\frac{\ell_*^u}{\ell_*}\Big),~~~
H_\ell=-\mathcal{X}_p
	\Big(\frac{\ell_*^\ell}{\ell_*}\Big).  \label{eq:Hk_poly}
\end{eqnarray}
%

\begin{figure*}
\includegraphics[width=170mm]{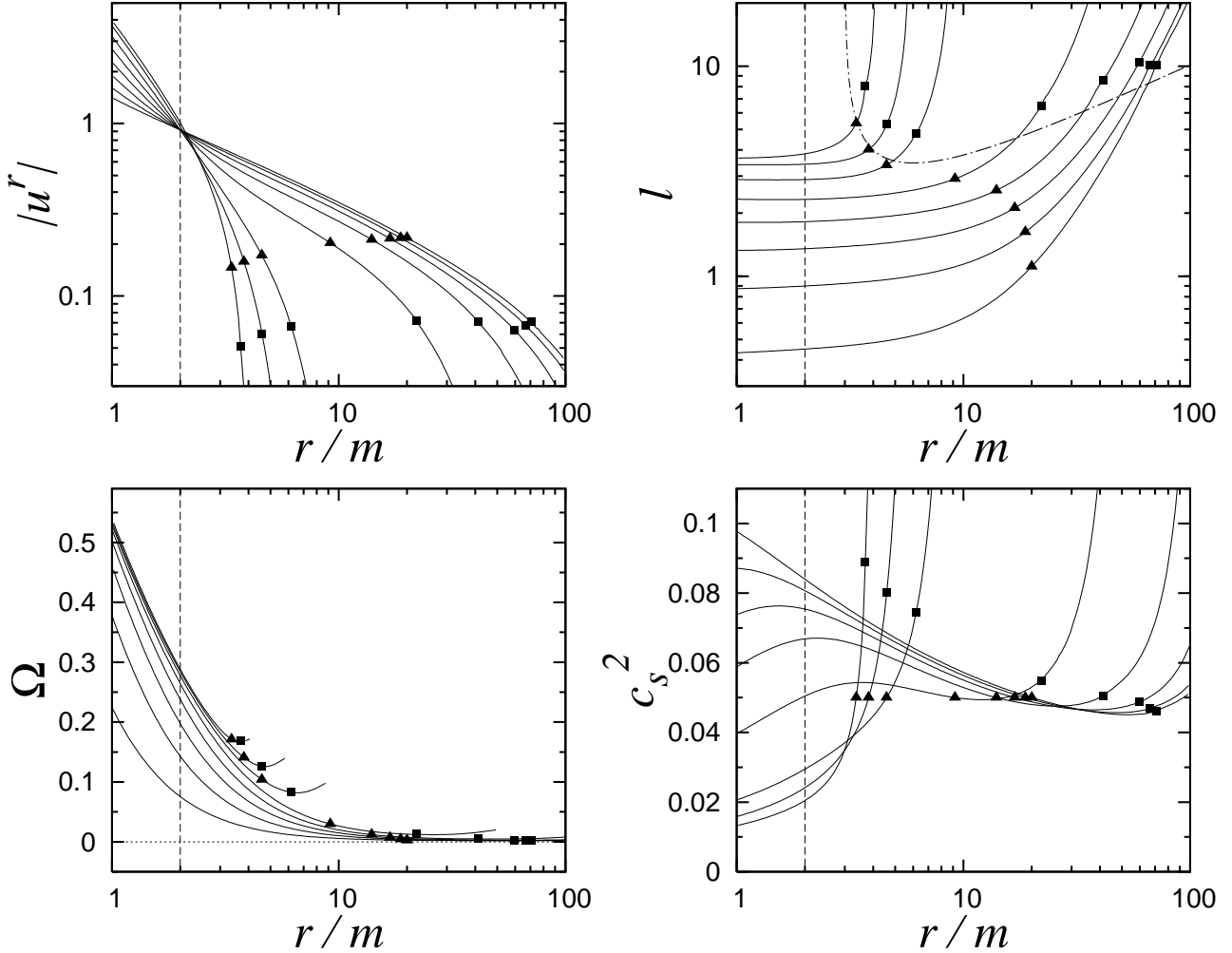}
\caption{
Sample transonic solutions for the viscous polytropic flows when $a/m=0$. 
We plot transonic solutions with the constant specific angular momentum 
$j=0.5$, 1.0, 1.5, 2.0, 2.5, 3.0, 3.5 and 3.7 (from right to left) 
with the critical sound speed $c_s^2=0.05$. 
The radius of the horizon is denoted by the dashed line.  
The sonic points and the viscous points are plotted by the filled triangles 
and squares, respectively.  %
}
\label{fig:PD_a00}
\end{figure*}

\begin{figure*}
\includegraphics[width=170mm,height=140mm]{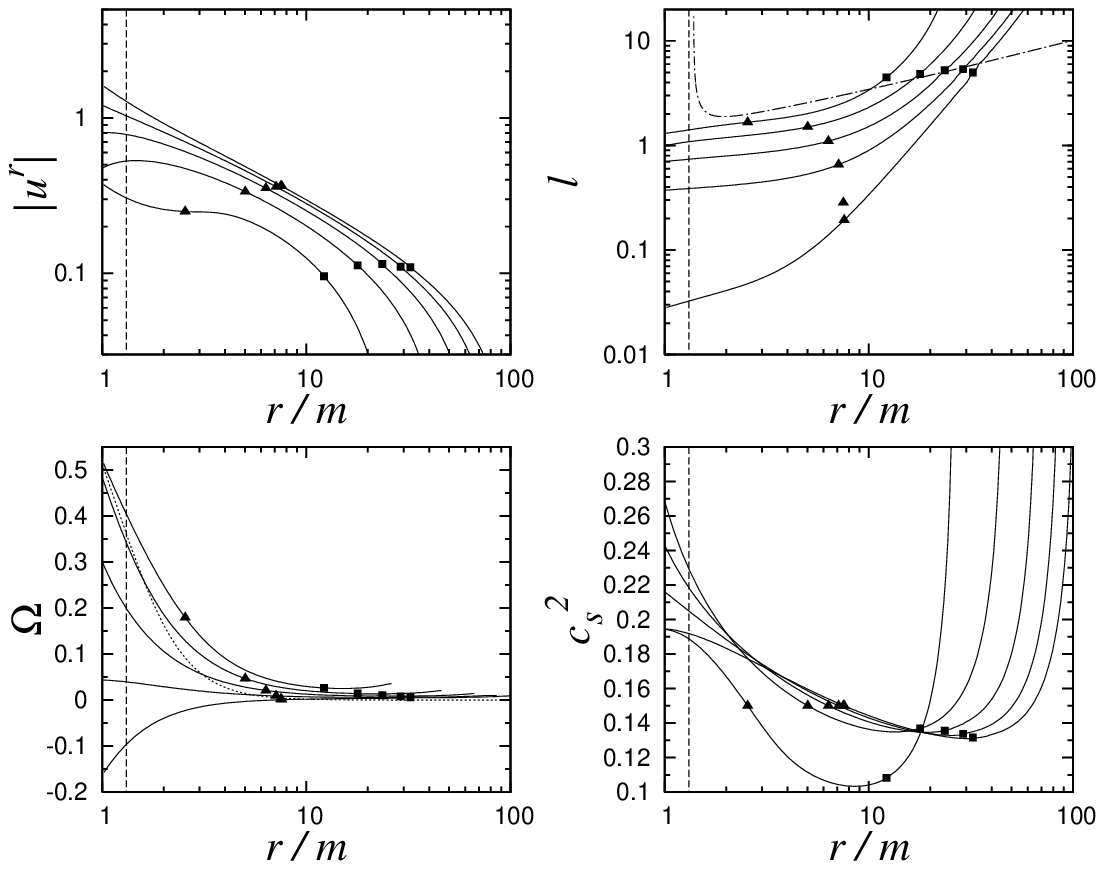}
\caption{
Sample transonic solutions for the viscous polytropic flows when $a/m=0.95$. 
We plot transonic solutions with the constant specific angular momentum 
$j=0.01$, 0.5, 1.0, 1.5 and 1.8 (from right to left) 
with the critical sound speed $c_s^2=0.15$. %
}
\label{fig:PD_a095}
\end{figure*}

By substituting Eqs. 
(\ref{eq:Pruell_poly}), 
(\ref{eq:eta_ruell_poly}) and (\ref{eq:Hk_poly})  
into Eqs. (\ref{eq:durdr_general}) and (\ref{eq:delldr_general}), the 
differential equations for $u^r$ and $\ell$ for the polytropic accretion flows 
are described as 
$du^r/dr=\mathcal{N}_s/\mathcal{D}_s$, 
$d\ell/dr=\mathcal{N}_v^r/\mathcal{D}_v+(\mathcal{N}_v^u/\mathcal{D}_v)du^r/dr$ 
which are the basic coupled differential equations to be solved. 
In this section, we calculate the ideal polytropic flows and the viscous polytropic 
flows. 
For the ideal polytropic flows, we assume $\eta\ell=$constant and only solve the differential 
equation for $u^r$, and the global transonic solutions do not have the viscous point.   
For the viscous polytropic flows, we solve the coupled differential equations of $u^r$ and 
$\ell$, and obtain the transonic solutions satisfying the boundary conditions at the sonic point 
and the viscous point.  
For all the viscous flows, we assume the alpha viscosity $\alpha_v=0.1$. 
The radius of the sonic point is determined in order to pass the viscous point. 
In the same way as the previous section, 
when calculating the numerical solutions, the rest-mass density $\rho_0$ is determined 
from $u^r$, $\ell$ and $c_s$ by using the mass conservation equation given by 
Eq. (\ref{eq:mass_cons2}). 
We show the sample solutions for the horizon-penetrating transonic solutions of 
the ideal polytropic flows for $a/m=0$ and $0.95$ in Fig. \ref{fig:PDideal_a00} 
and Fig. \ref{fig:PDideal_a095}, respectively. 
The sonic points are plotted by the filled triangles, and the radius of the horizon is 
plotted by the dashed lines. 
The transonic solutions for the viscous polytropic flows for $a/m=0$ and $0.95$ 
are given in Fig. \ref{fig:PD_a00} and Fig. \ref{fig:PD_a095}, respectively. 
For the viscous flows, the sonic points and the viscous points are plotted by 
the filled triangles and squares, respectively. 
All the viscous polytropic solutions presented here become super-Keplerian flows 
in the outer region, and the sound speed diverged. 
The outer region of these solutions correspond to the thick disk solutions. 
%


\subsection{Application 3 : ADAFs with Relativistic Equation of State}
\label{sec:ADAF}
Here, we calculate the transonic solutions for the ADAF with relativistic equation of state. 
where the energy equation given by Eq. (\ref{eq:energy}) is required 
in order to close the coupled differential equations. 
For the Boyer-Lindquist coordinate which have the coordinate singularity at the horizon, 
\cite{gp98} and \cite{pg98} solve the transonic 
solutions by using the causal viscosity prescription. 
%

\begin{figure*}
\includegraphics[width=170mm]{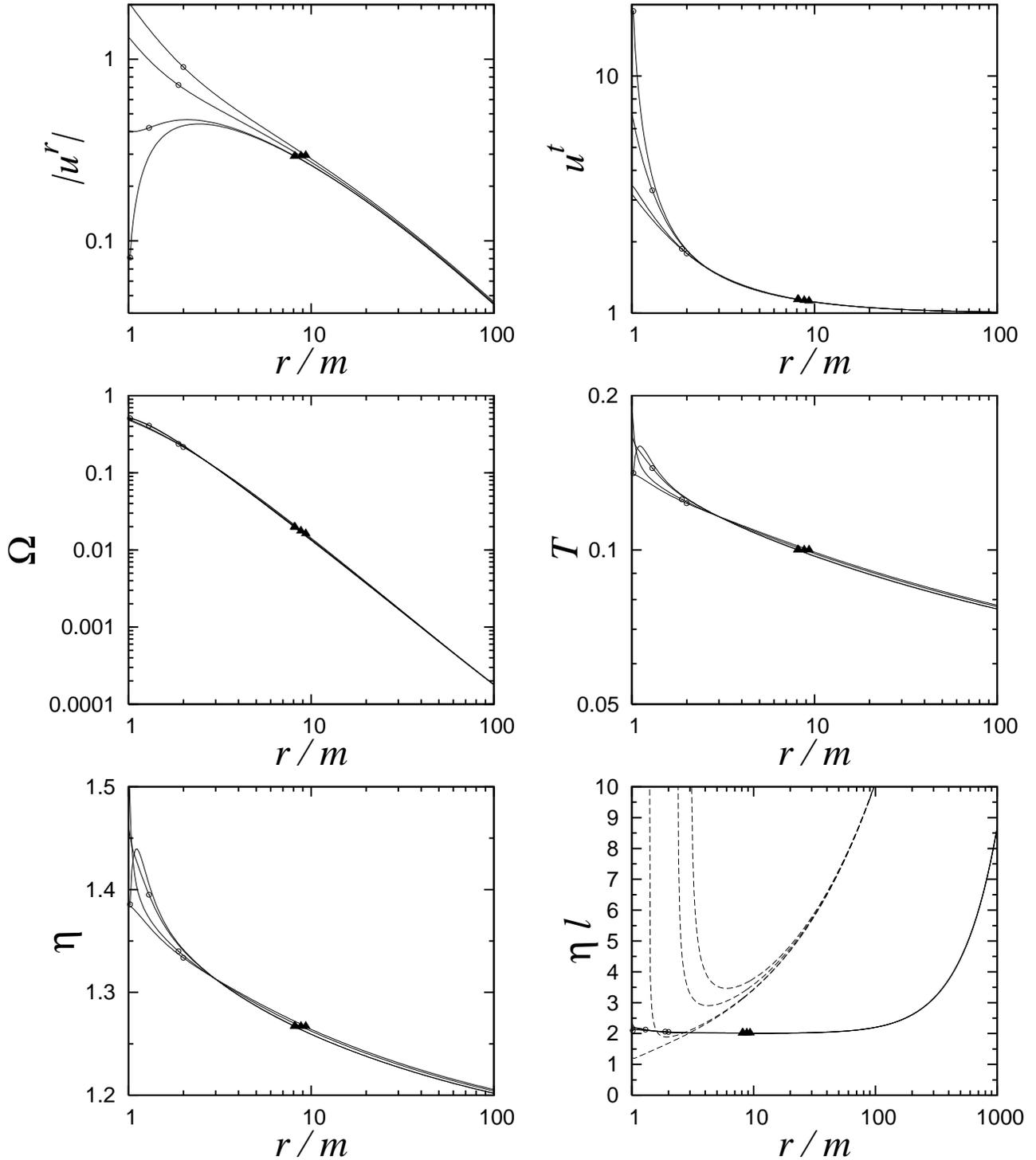}
\caption{
Sample numerical solutions for transonic accretion flows of ADAFs 
with relativistic equation of state 
when $a/m=0.0$, 0.5, 0.95 and 0.99999 with $T_s=0.1$. 
We plot the four-velocity components, $u^r$ and $u^t$, 
the angular velocity $\Omega$, 
the dimensionless temperature $T$, the relativistic enthalpy $\eta$ 
and $\eta\ell$. 
The solutions are calculated so as to satisfy zero shear stress 
at the horizon. 
The positions of the horizons and the sonic points are denoted 
by the blank circles and the filled triangles, respectively.
The dashed lines in the panel for $\eta\ell$ ({\it bottom right}) are 
the angular momentum for the Keplerian motion when 
$a/m=0.0$, 0.5, 0.95 and 0.99999 (right to left). 
Here, we assume the alpha viscosity parameter 0.01 
and type A causal viscosity. 
}
\label{fig:ADAF}
\end{figure*}

%
%
%
Here, we use the equation of state same as \cite{gp98}. 
The pressure $p$ and the internal energy $u$ is given by the rest-mass density $\rho_0$ 
and the temperature $T$ as \citep{c39,cg68}
\begin{eqnarray}
p&=&\rho_0 T, \label{eq:p_ADAF}\\
u&=&\rho_0 \left[3T +\frac{K_1(1/T)}{K_2(1/T)}\right],\label{eq:u00_ADAF}
\end{eqnarray}
%
where $K_n$'s are the modified Bessel functions of the second kind of order $n$. 
The internal energy $u$ is well fitted by the function as \citep{gp98} 
$u=\rho_0 T g(T)$ 
where $g(T)=(45T^2+45T+12)/(15T^2+20T+8)$. 
The relativistic enthalpy $\eta$ and the sound velocity $c_s$ 
become a function of the temperature $T$ as, 
\begin{equation}
\eta=1+T\left[1+g(T)\right],~~~~~
c_s^2=\frac{T}{1+T[1+g(T)]}. \label{eq:eta_as_ADAF}
\end{equation} 
Here, $\eta$ and $c_s$ are functions of temperature, $T$. 
%
%
%
%
By using these equations, we can obtain 
\begin{equation}
P_\rho=1,~~~
P_T=1,~~~
U_\rho=1,~~~
U_T=1+\frac{d\ln g(T)}{d\ln T},~~~
\eta_\rho=0,~~~
\eta_T=\frac{1}{\rho_0 \eta}\left(1+u+u\frac{d\ln g(T)}{d\ln T}\right). \label{eq:PUeta_ADAF}
\end{equation}
On the other hand, by using the assumption $q^-_{\rm rad}=0$, 
we can calculate the coefficients 
$q^\pm_k$ ($k=r$, $u$ and $\ell$) 
as 
\begin{equation}
q^\pm_k=-2S\sigma_k. 
\end{equation}
By substituting equations given by Eq. (\ref{eq:PUeta_ADAF}) into 
Eqs. (\ref{eq:T_k}), (\ref{eq:P_k}) and (\ref{eq:H_k}), 
we can calculate the coefficients $P_k$, $H_k$, $\eta_k$ and $T_k$ ($k=r$, $u$ and $\ell$). 
Now, it is noted that the relativistic enthalpy $\eta$ is also calculated 
by the equation in Eq. (\ref{eq:eta_as_ADAF}). 
Then, the derivatives $du^r/dr$, $d\ell/dr$ and $dT/dr$ are obtained and numerically 
solved in order to calculate the transonic solutions for the adiabatic accretion disks. 
In Fig. \ref{fig:ADAF}, we show the sample numerical transonic solutions for 
the ADAF with the relativistic equation of state. 
Sample numerical solutions are calculated for  
$a/m=0.0$, 0.5, 0.95 and 0.99999 with $T_s=0.1$, and 
we plot the four-velocity components, $u^r$ and $u^t$, the angular velocity $\Omega$, 
the dimensionless temperature $T$, the relativistic enthalpy $\eta$ and $\eta\ell$. 
The solutions are calculated so as to satisfy zero shear stress at the horizon. 
The positions of the horizons and the sonic points are denoted 
by the blank circles and the filled triangles, respectively.
The dashed lines in the panel for $\eta\ell$ ({\it bottom right}) are 
the angular momentum for the Keplerian motion when 
$a/m=0.0$, 0.5, 0.95 and 0.99999 (right to left).
All solutions pass the event horizon smoothly and have nearly same flow patterns 
in the outer region. 
As denoted in the previous section, the accretion flow is plunging into 
the black hole with the angular velocity which is different 
from the angular velocity of the black hole's rotation at the event horizon. 
While the accretion flow have the smaller values of $|u^r|$ 
for larger values of black hole spins, the gamma factor $\gamma=\alpha u^t$ become 
larger for the larger values of the black hole spins. 
This is because of the effects of the black hole rotation enhances the rotational 
velocity of the accretion flow, i.e., 
for larger black hole spins the accretion flows have the larger values of 
the angular three velocity. 
The relativistic enthalpy $\eta=1+(u+p)/\rho_0$ of the hot accretion flow 
is in general larger than unity near the horizon. 
In the sample solutions for $a/m=0.95$ and 0.99999, the accretion flows are sub-Keplerian 
in all the region. 
On the other hand, for $a/m=0.5$, the middle part of the accretion flow is super-Keplerian. 
The similar feature is also pointed out by \cite{pa97}. 
%

\subsection{Application 4 : Adiabatic Accretion Disk and Standard Accretion Disk}
\label{sec:AD}
\begin{figure*}
\includegraphics[width=170mm]{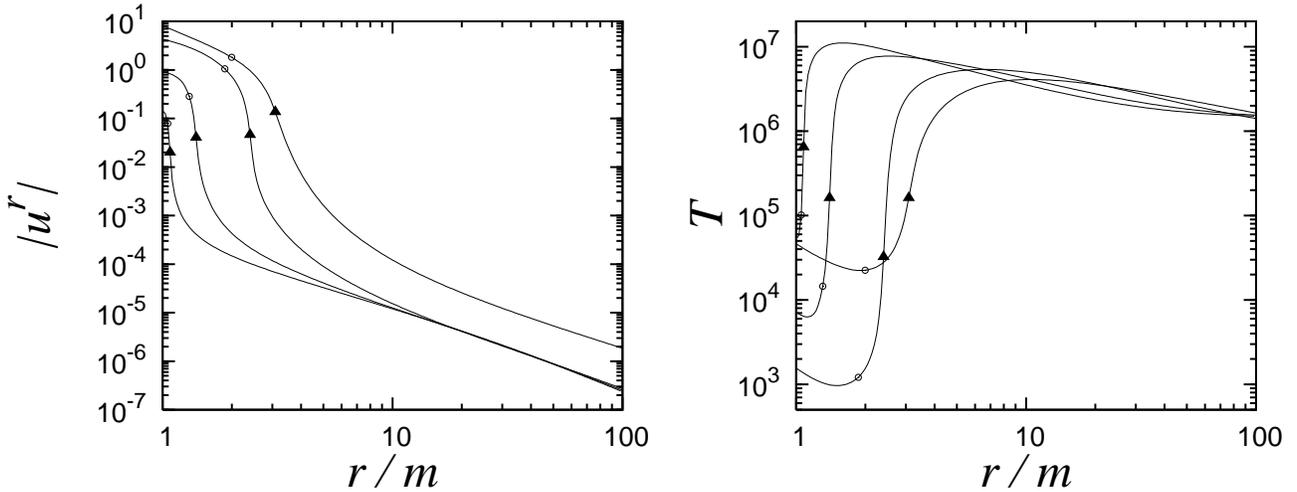}
\caption{
Sample numerical solutions for the adiabatic accretion disks when 
$a/m=0.0$, 0.5, 0.95 and 0.998 with dimensionless temperature 
$T_s=0.05$, 0.01, 0.05 and 0.2, respectively.  
The black hole mass $M_{\rm BH}=10M_\odot$ and 
the mass accretion rate $\dot{M}=10 \dot{M}_{\rm Edd}$ are assumed. 
The radial component of four velocity $u^r$ and the dimensional temperature 
$T$ [K] are plotted. 
The positions of the horizons and the sonic points are denoted 
by the blank circles and filled triangles, respectively.  
The solutions are calculated so as to satisfy zero shear stress at the horizon. 
Here, the alpha viscosity parameter 0.01 and type A viscosity are assumed. 
}
\label{fig:AD}
\end{figure*}

In this section, we calculate the accretion flows where the viscous heating rate 
is balanced with the radiative cooling rate, $q_{\rm vis}^+=q_{\rm rad}^-$. 
This assumption is usually used in the calculations of the standard accretion disks. 
From this assumption and the energy equation given 
by Eq. (\ref{eq:energy2}), we can show that the entropy change 
of the accretion flow is zero, i.e. $dS=0$. 
So, here, we call the disk with this assumption as adiabatic accretion disk. 
Here, we calculate the transonic solutions for the adiabatic accretion disks
whose mass accretion rate is near or beyond the Eddington mass accretion rate. 
For the supercritical accretion flow the contribution from the radiation pressure of photons 
can not be neglected. 
When the specific heat at the constant volume, $c_V$, is independent of the temperature, 
the pressure $p$ and the internal energy $u$ for the flows containing gas and radiation 
are given as \citep{c39}, 
%
\begin{eqnarray}
p&=&p_g+p_r, \label{eq:p_SCD}\\
u&=&\frac{1}{\gamma-1}p_g +3p_r, \label{eq:u_SCD} 
\end{eqnarray}
%
where $p_g$ and $p_r$ are the gas pressure and the radiation pressure, respectively, described as 
\begin{equation}
p_g=\frac{k_B}{\mu m_H}\rho_0 T,~~~~~p_r=\frac{1}{3}a_{\rm rad}T^4. 
\end{equation}
Here, $\gamma$ is the ratio of the specific heats, $k_B$ is the Boltzmann constant, $\mu$ is 
the mean molecular weight, $m_H$ is the Hydrogen mass, $a_{\rm rad}$ is the radiation constant, 
and we use the dimensional representation of the pressures. 
%
%
%
%
These assumptions are sometimes used in the past studies for the standard accretion disk 
(e.g. Shakura \& Sunyaev 1973, Novikov \& Thorne 1974, 
Page \& Thorne 1974, Matsumoto et al. 1984).  
%
For flows with $dS=0$, the energy equation, 
$u^r\rho_0T(dS/dr)=q_{\rm vis}^+-q_{\rm rad}^-(=0)$,  
can be given by using the generalized 
adiabatic exponents as 
\begin{equation}
\frac{u^r}{\Gamma_3-1}\left(\frac{dp}{dr}-\Gamma_1\frac{p}{\rho_0}\frac{d\rho_0}{dr}\right)
=q_{\rm vis}^+-q_{\rm rad}^-(=0), 
\label{eq:energy_AD}
\end{equation}
where 
\begin{eqnarray}
\Gamma_1&=&\beta+\frac{(4-3\beta)^2(\gamma-1)}{\beta+12(\gamma-1)(1-\beta)},\\
\Gamma_3&=&1+\frac{\Gamma_1-\beta}{4-3\beta}.
\end{eqnarray}
Here, $\beta$ is the ratio of the gas pressure to the total pressure, i.e. 
$\beta\equiv p_g/p$. 
In the calculations of the transonic flows, we use the dimensionless pressure, 
internal energy, rest-mass density and the temperature. 
By using the dimensionless prescription, the pressure and the internal energy 
can be calculated as $p=p_g+p_r$ and $u=p_g/(\gamma-1)+3p_r$ where $p_g=\rho_0 T$ and 
$p_r=T^4/3$ where the temperature is normalized by $m_p c^2/k_B$, 
the rest-mass density is firstly normalized by $(a_{\rm rad}/c^2)(m_p c^2/k_B)^4$ 
to the dimensionless quantities and is secondly normalized so as to satisfy $\dot{M}=1$. 
Here, $m_p$ is the proton mass and $c$ is the speed of light. 
Since the left-hand-side of Eq. (\ref{eq:energy_AD}) derived from the condition $dS=0$ 
is equivalent to that of Eq. (\ref{eq:energy2}), we have the relations 
\begin{eqnarray}
\left(\frac{\partial u}{\partial T}\right)_{\rho_0}
	&=&\frac{1}{\Gamma_3-1}\left(\frac{\partial p}{\partial T}\right)_{\rho_0}, 
\label{eq:dudT_AD}\\
\left(\frac{\partial u}{\partial \rho_0}\right)_{T}
	&=&\frac{1}{\Gamma_3-1}\left[
	\left(\frac{\partial p}{\partial \rho_0}\right)_{T}
	-\Gamma_1\frac{p}{\rho_0}
	\right]+\frac{u+p}{\rho_0}. 
\label{eq:dudrho_AD}
\end{eqnarray}
From the equation of state and Eqs. (\ref{eq:dudT_AD}) and (\ref{eq:dudrho_AD}), we have 
\begin{equation}
P_\rho=\beta,~~
P_T=4-3\beta,~~
U_\rho=-\frac{p}{u}\left[2(4-3\beta)+\frac{\beta}{\gamma-1}\right],~~
U_T=\frac{p}{u}\left(\frac{4-3\beta}{\Gamma_3-1}\right),~~  
\label{eq:PUeta_AD}
\end{equation}
%
where $u/p=\beta/(\gamma-1)+3(1-\beta)$. 
By substituting equations in Eq. (\ref{eq:PUeta_AD}) into 
Eqs. (\ref{eq:T_k}), (\ref{eq:P_k}) and (\ref{eq:H_k}), 
we can calculate the coefficients $P_k$, $H_k$, $\eta_k$ and $T_k$ ($k=r$, $u$ and $\ell$). 
Then, the derivatives $du^r/dr$, $d\ell/dr$ and $dT/dr$ are obtained and numerically 
solved in order to calculate the transonic solutions for the adiabatic accretion disks. 
In Fig. \ref{fig:AD}, we show the sample transonic solutions of the adiabatic accretion 
disks with the black hole mass $M_{\rm BH}=10M_\odot$ and 
the mass accretion rate $\dot{M}=10 \dot{M}_{\rm Edd}$ where 
$\dot{M}_{\rm Edd}=1.4\times 10^{17}(M_{\rm BH}/M_\odot)$ [g s$^{-1}$]  
is the Eddington mass accretion rate. 
Sample solutions are calculated for  
$a/m=0.0$, 0.5, 0.95 and 0.998 with dimensionless temperature 
$T_s=0.05$, 0.01, 0.05 and 0.2, respectively.  
In Fig. \ref{fig:AD}, 
the radial component of four velocity $u^r$ ({\it left panel})
and the dimensional temperature $T$ [K] ({\it right panel}) are plotted. 
The positions of the horizons and the sonic points are denoted 
by the blank circles and filled triangles, respectively.  
The solutions are calculated so as to satisfy zero shear stress at the horizon. 
The temperatures of all solutions become smaller by several order inside 
the marginally stable orbit. 
This feature is same as the standard accretion disks. 
On the other hand, inside the marginally stable orbit, the absolute value of 
the radial component of the four velocity become large. 
This is because the accretion flow approach the free fall motion with angular momentum 
in this region and plunge into the black hole horizon with high value of 
the gamma factor.  
%

\subsection{Application 5: Supercritical Accretion Disk with Photon-trapping Effects}
\label{sec:SCD}
Based on the formulation described above, we also calculate the simplest version of the 
transonic solutions of the slim disk of the supercritical accretion flow where the mass 
accretion rate is larger than the super-Eddington mass accretion rate. 
In general, for the supercritical accretion disks, the assumption of adiabatic changes, $dS=0$, 
is not valid because in the vicinity of a black hole the pressure gradient enhance 
the radial velocity of the flow and the advection term in the energy equation can not be neglected. 
In such case, the energy equation with the effects of advection cooling, the radiation cooling and the viscous heating should be solved. 
In addition, near the horizon, photons are trapped within matters and can not 
escape from the accretion flow, and the flow become advection dominated state.  
The past studies actually solve the transonic solutions of the supercritical 
accretion flow (e.g. Watarai et al. 2001, Watarai \& Mineshige 2001, Shimura \& Manmoto 2003) 
by assuming that the specific heat at the constant volume, $c_V$, is independent of 
the temperature.  
For supercritical accretion flow, 
the heat inertia can not be neglected \citep{b98} and the photon-trapping effects near the 
horizon is also important. 
For the Boyer-Lindquist coordinate which have the coordinate singularity at the horizon, 
\cite{b98} and \cite{sm03} solve the transonic solutions 
based on the acausal viscosity prescription. 
From here, we just show the formulation of the transonic solutions for the supercritical 
accretion flows with effects of the heat inertia and the photon trapping with the general 
form of the specific heat at the constant volume, i.e., here, we do not assume that 
the specific heat at the constant volume, $c_V$, is independent of the temperature. 
For such flows, the internal energy $u$ is calculated as 
\begin{equation}
u=p_g g(T) + 3p_r, 
\end{equation}
where $g(T)$ is the same function used in the calculations of the transonic solutions 
of ADAFs with relativistic equation of state described in the previous section. 
First, we roughly estimate the photon trapping effects around black holes. 
In the optically thick region for photons in the supercritical accretion flows, 
the photons can escape from the disk surface after the diffusive processes in the disk. 
This holds only when the radiative diffusion timescale is shorter than the 
accretion timescale (see e.g. Katz 1977, Begelman 1978, Ohsuga et al. 2002, Kawaguchi 2003). 
The diffusion velocity is roughly calculated as $v_{\rm diff}\sim c/(3\tau)$. 
Then, the diffusion timescale of photons produced at equatorial plane 
is written as $t_{\rm diff}=H/v_{\rm diff}$. 
On the other hand, the accretion time scale is $t_{\rm acc}=r/(-u^r)$. 
When $t_{\rm diff}> t_{\rm acc}$, 
photons are trapped in the accretion disk and plunged into black hole 
without escaping from the disk surface. 
From the condition $t_{\rm diff}> t_{\rm acc}$, we can derive 
the photon-trapping radius, $r_{\rm trap}^*$, 
within which the parts of photons begin to be trapped as 
$r_{\rm trap}^*=[3\bar{\kappa}/(2\pi c)]H_\theta\dot{M}$  
where $\bar{\kappa}$ is the mean opacity for photons. 
The radiation term including effects of the 
photon-trapping, the electron scattering and the free-free absorption is calculated as 
$q_{\rm rad}^-=f_{\rm trap}q_{{\rm rad},0}^-$ where 
$q^-_{{\rm rad},0}$ is the cooling rates when no effects of 
photon trapping. 
Now, $f_{\rm trap}=1$ means no photon trapping and 
$f_{\rm trap}=0$ means complete photon trapping. 
In this study, 
we assume $f_{\rm trap}=1$ for $r>r_{\rm trap}$, but 
$f_{\rm trap}=0$ for $r\leq r_{\rm trap}$, for simplicity. 
The radiative cooling term, $q_{{\rm rad},0}^-$ [erg cm$^{-3}$ s$^{-1}$],  
without the photon trapping is calculated as 
$q_{{\rm rad},0}^-=F^-/(2rH_\theta)$ 
%
%
where $F^-$ [erg cm$^{-2}$ s$^{-1}$] is 
the energy loss rate from the disk surface which is calculated by using 
the Rosseland approximation as 
$F^-=(16\sigma_{\rm SB} T^4)/(3\bar{\kappa}\rho_0 rH_\theta)$ in the outside region, 
and $F^-$ is calculated by the free-free emission inside region \citep{b98}. 
%
%
Here, $\sigma_{\rm SB}$ is the Stephan-Boltzmann constant. 
We can calculate the coefficients $q^\pm_k$ ($k=r$, $u$ and $\ell$) as 
\begin{equation}
q^\pm_r=-2S\sigma_r-q_{\rm rad}^-,~
q^\pm_u=-2S\sigma_u,~
q^\pm_\ell=-2S\sigma_\ell. \label{eq:qk_SCD}
\end{equation}
On the other hand, from the EOS given by Eqs. (\ref{eq:p_SCD}) and (\ref{eq:u_SCD}), 
we obtain 
\begin{equation}
P_\rho=\beta,~~
P_T=4-3\beta,~~
U_\rho=\frac{\beta g(T)p}{u},~~
U_T=\frac{p}{u}\left[\beta\left(g+T\frac{d g}{d T}\right)+12(1-\beta)\right],
\label{eq:PUeta_SCD}
\end{equation}
where $u/p=\beta g(T)+3(1-\beta)$ and $dg/dt=15(15T^2+24T+8)/(15T^2+20T+8)^2$. 
By substituting Eqs. (\ref{eq:PUeta_SCD}) and (\ref{eq:qk_SCD}) into 
Eqs. (\ref{eq:T_k}), (\ref{eq:P_k}) and (\ref{eq:H_k}),  
we can calculate the coefficients $P_k$, $H_k$, $\eta_k$ and $T_k$ ($k=r$, $u$ and $\ell$). 
Then, the derivatives $du^r/dr$, $d\ell/dr$ and $dT/dr$ are obtained and numerically 
solved in order to calculate the transonic solutions for the supercritical accretion disks with 
effects of advection cooling. 
%

\section{Concluding Remarks}
%
%
\begin{figure*}
\includegraphics[width=100mm]{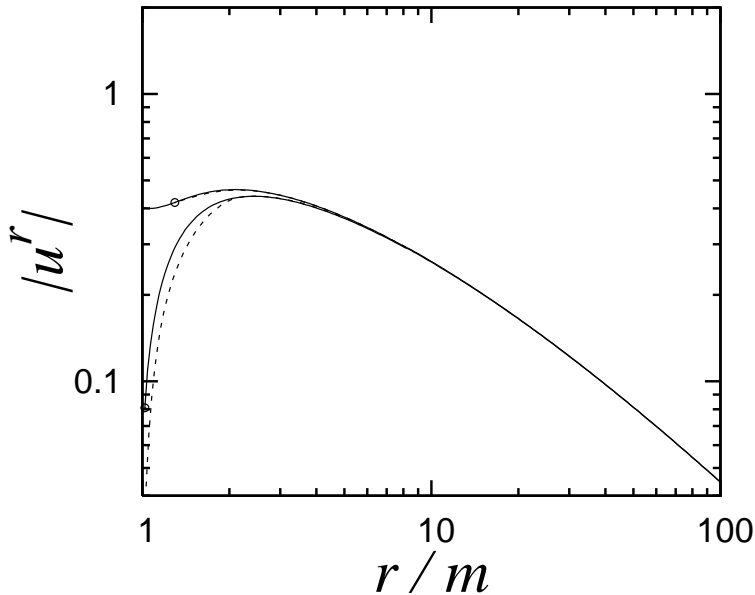}
\caption{
Transonic solutions of ADAF calculated by using the Kerr-Schild 
coordinate ({\it silid lines}) and 
the Boyer-Lindquist coordinate ({\it dashed lines})
for $a/m=0.95$ and 0.99999. 
The parameters are same as Fig. \ref{fig:ADAF}.  
The blank circles show the position of the horizon. 
}
\label{fig:VisInv}
\end{figure*}
%
%
Before summing up the results of this study, 
it may be better to note the arguments about the causality 
of the viscous flows at the event horizon. 
As already pointed out by Popham \& Gammie (1998), 
the formalism of the causal 
viscosity prescription give the solutions with the finite values 
of the outward energy flux and 
the outward angular momentum at the horizon, and they stated 
that this property does not suggest 
the causality violation (Popham \& Gammie 1998). 
If the viscosity is described by the fluctuations of 
the Maxwell stress and/or the Reynolds 
stress from the mean values such as the angular momentum transport by 
the magnetorotational instability (MRI), these fluctuating parts 
should be correctly treated 
in the general relativistic point of view which may require 
the extended causal thermodynamics 
such as the Israel-Stewart theory \citep{is79} 
where the causality violating infinite signal speeds are eliminated. 
The hydrodynamical equations based on such theory are 
formulated by \cite{pa98}. 
If the extended causal thermodynamics for the magnetohydrodynamical 
flows can be used, 
the problems of the causality of the viscous flows at 
the event horizon will be clearly 
resolved in the future. 
The another limitation of the viscosity prescription used in this 
paper is shown in Fig. \ref{fig:VisInv}. 
In this figure, we show 
the transonic solutions of ADAF calculated by using the Kerr-Schild 
coordinate ({\it silid lines}) and 
the Boyer-Lindquist coordinate ({\it dashed lines})
for $a/m=0.95$ and 0.99999. 
The parameters are same as Fig. \ref{fig:ADAF}.  
For the case of $a/m=0.95$, the two solutions have almost same results. 
For the cases of lower spin parameters than $a/m=0.95$, 
the same solutions by using the two coordinate are obtained. 
However, in the case of $a/m=0.99999$ shown in Fig. \ref{fig:VisInv}, 
the two results are not exactly same. 
This feature show that the viscosity prescription used in this paper  
is not perfectly coordinate invariant. 
When we introduce the causality limited viscosity, 
we use some special frame in order to evaluate the physical quantities 
such as the shear stress $S$ or the factor $f_c$. 
However, there is no guarantee that these physical quantities 
introduced in the specific reference frame have the invariant 
feature with respect to the coordinate transformation or 
the frame transformation. 
If we can define and introduce the invariant viscosity, 
the accretion flows calculated by the two procedures produce 
the same results such as the ideal flows. 
So, the viscous flows presented in the previous sections are 
not very exactly reproduced from the calculations 
based on the Boyer-Lindquist coordinate such as 
the past studies.  
The invariant viscosity prescription will be made 
by the extended causal thermodynamics stated above. 
From here, we give the conclusions of the present study. 
In the present study, 
we give the basic equations and the calculation method 
for the horizon-penetrating transonic accretion disks or 
flows in the equatorial plane written by the Kerr-Schild coordinate 
where there is no 
coordinate singularity at the event horizon. 
Based on these formalism, we calculate the transonic solutions of 
these types of the 
accretion flow models from the outer region to inside region of the 
event horizon; 
the ideal isothermal flows, the ideal and the viscous polytropic flows, and 
the advection dominated accretion flows (ADAFs) 
with the relativistic equation of state, the adiabatic accretion disks, 
the standard accretion disks and the supercritical accretion disks.   
In this study, we use two types of the causal viscosity prescriptions. 
One is the simple treatment of the kinematic viscosity and the 
other is based 
on the shear stress measure in the FRF. 
When we use the causal viscosity prescription based on the shear 
stress measured in the FRF, 
the boundary condition for the transonic accretion flows is also given  
at the viscous point where the accreting radial velocity is nearly equal 
to the viscous diffusion velocity. 
By using the causal viscosity prescription, the ADAF transonic solutions are 
firstly obtained by \cite{gp98} for the general relativistic flows 
and later for the 
pseudo-Newtonian flows \citep{t07}.  
Based on the solutions obtained in the present study, 
we calculate the physical values for the transonic solutions of the 
these disks 
around the rotating black hole just on the event horizon 
and inside the horizon. 
These solutions are obtained for both non-rotating and rotating black holes. 
In general, the accretion flows calculated by using 
the Kerr-Schild coordinate 
plunge into black hole with finite three velocity 
smaller than the speed of light even at the event horizon or 
inside the horizon, 
and the angular velocities at the horizon are  different from the angular 
velocity of the frame-dragging due to the black hole's rotation. 
These features are different from the results obtained 
by using the Boyer-Lindquist 
coordinate with the coordinate singularity at the horizon. 
By using the formalism presented in the present study and 
adding the required physics, 
we can basically calculate the another types of 
more realistic accretion flows including 
the radiatively inefficient accretion flows (RIAF) in the galactic center, 
the supercritical accretion disks which is sometimes assumed 
in the center of the 
black-hole X-ray binaries or Syfert galaxies, and 
the hypercritical accretion flows or 
the neutrino-dominated accretion flows (NDAFs) in gamma-ray burst. 
Also, by using the formulations and the calculation methods 
in this study, the accretion flows 
inside the black hole satisfying the boundary conditions outside 
the black hole can be 
calculated. 
Although the accretion flow structure inside the event horizon 
can not be seen directly 
by the observer outside the event horizon, 
by combining the constraints obtained by the future observations 
for the regions 
just outside the event horizon of the black hole candidate such 
as the massive black hole at the galactic center 
with the theoretical calculations such as this study or more sophysticated 
study inside the black hole, 
we can know the accretion flow structure inside the event horizon of 
the black hole in the real world in the future.  
Especially, by the near-future observations by 
the radio interferometer such as 
e.g. VSOP-2 for radio (Hirabayashi et al. 2005), 
MAXIM for soft X-ray (see MAXIM web page: http://maxim.gsfc.nasa.gov/), 
the direct-mapping of the black hole shadow in the RIAF 
in the galactic center will be performed 
(e.g. Falcke et al. 2000; Melia \& Falcke 2001 
for review, Melia 2003a, 2003b and references therein) 
and give the information of the strong-gravity region 
as the resolved images around the shadow. 
It is known that such images will give the physical information 
of the black hole 
itself or the accretion flows in the strong-gravity region 
(e.g. Cunningham \& Bardeen 1972, 1973, Bardeen 1973, Takahashi 2004, 2005, 
Broderick \& Loeb 2005, 2006, Broderick \& Narayan 2006, 
Zakharov et al. 2005, Yuan et al. 2006) . 
However, so far, the image of the black hole shadows calculated by using 
the general relativistic transonic flows of the RIAF have 
not been performed.   
Our calculations presented in this study can be also applied 
to such calculations 
with the radiation mechanisms and the required physics.  
%

\section*{Acknowledgments}
The author is grateful to 
Professors Y. Eriguchi and S. Mineshige 
for their continuous encouragements,   
and K. Watarai, S. Mineshige, K. Ohsuga, M. Takahashi, K. Nakao, 
S. Nagataki, J. Fukue, R. Matsumoto, N. Kawanaka, Y. Sekiguchi, 
M. Shibata, T. Yamamoto and A. Yoshinaga 
for useful discussion and comments. 
The author also thanks the anonymous referee for helpful 
and useful comments. 
This research was partially supported by the Ministry of Education,
Culture, Sports, Science and Technology, Grant-in-Aid for 
Japan Society for the Promotion of Science (JSPS) Fellows (17010519).

\appendix

\section{Metric Components}
\label{app:metric}
The components of $g_{\mu\nu}$ and $g^{\mu\nu}$ are calculated as 
\begin{equation}
g_{\mu\nu}=
\left(
	\begin{array}{cc}
		-\alpha^2+\beta^k \beta_k & \beta_i \\
                \beta_i & \gamma_{ij}
	\end{array}
\right),
\end{equation}
and 
\begin{equation}
g^{\mu\nu}=
\left(
	\begin{array}{cc}
        -1/\alpha^2 & \beta^i/\alpha^2 \\
		\beta^i/\alpha^2 & \gamma^{ij}-\beta^i\beta^j/\alpha^2
	\end{array}
\right).   
\end{equation}
%
The explicit forms of the non-zero components of 
$g_{\mu\nu}$ and $g^{\mu\nu}$ are calculated as 
%
\begin{eqnarray}
g_{tt}&=& -\left(1-\frac{2mr}{\Sigma}\right),~~~
g_{tr}=g_{rt}= \frac{2mr}{\Sigma},~~~
g_{t\phi}=g_{\phi t}= -\frac{2mar\sin^2\theta}{\Sigma},~~~
g_{rr}= 1+\frac{2mr}{\Sigma},\nonumber\\
g_{r\phi}&=&g_{\phi r}= -a\sin^2\theta\left(1+\frac{2mr}{\Sigma}\right),~~~
g_{\theta\theta}= \Sigma,~~~
g_{\phi\phi} = \frac{A\sin^2\theta}{\Sigma},
\end{eqnarray}
%
and 
%
\begin{eqnarray}
g^{tt}&=& -\left(1+\frac{2mr}{\Sigma}\right),~~~
g^{tr}=g^{rt}= \frac{2mr}{\Sigma},~~~
g^{rr}= \frac{\Delta}{\Sigma},~~~
g^{r\phi}=g^{\phi r}= \frac{a}{\Sigma},~~~
g^{\theta\theta}= \frac{1}{\Sigma},~~~
g^{\phi\phi} = \frac{1}{\Sigma\sin^2\theta}, 
\end{eqnarray}
where nonzero components of $\beta_i$ and $\gamma^{ij}$ are 
%
\begin{eqnarray}
\beta_r&=& g_{tr}=\frac{2mr}{\Sigma},~~~
\beta_\phi= g_{t\phi}=-\frac{2mar\sin^2\theta}{\Sigma},~~~
\gamma^{rr}=\frac{A}{\Sigma(\Sigma+2mr)},~~~
\gamma^{r\phi}=\gamma^{\phi r}=\frac{a}{\Sigma},~~~
\gamma^{\theta\theta}=\frac{1}{\Sigma},~~~
\gamma^{\phi\phi}=\frac{1}{\Sigma\sin^2\theta}. \nonumber
\end{eqnarray}
%
Here, we use $\beta_i=\gamma_{ij}\beta^j$ and 
$\gamma^{ik}\gamma_{kj}=\delta^i_j$ where $\delta^i_j$ 
represents the Kronekker delta.

\section{Congruences for the Observer Dragging with 
the Black Hole's Rotation}
\label{app:singular_frame}
Here, we show that 
the congruences for the observer rotating with the angular velocity of 
the frame-dragging written by the Kerr-Schild coordinate 
have the singularity at the event horizon. 
We consider 
the observer rotating with the angular velocity, $\omega=2mar/A$, which is the angular velocity 
of the frame dragging due to the black hole's rotation.   
For such observer, by using the normalization condition $u^\mu u_\mu=-1$, 
the contravariant components of the four velocity are described as 
%
\begin{equation}
u^{t}= \left(\frac{\Delta\Sigma}{A}\right)^{1/2},~~~
u^{r}= u^{\theta}=0,~~~
u^{\phi}=\omega u^{t}.   
\end{equation}
Since $\Delta<0$, $\Sigma>0$ and $A>0$ within the event horizon, 
the component $u^t$ become imaginary. 
Then, in this study, the congruences for the observer moving with 
are not used when the transformation of the physical quantities between the KSF and the FRF.

\section{Transformation between the Kerr-Schild frame and the Fluid's Rest Frame by Tetrads}
\label{app:tetrads}
First, we give the tetrad components connecting the KSF and 
the LNRF made by the congruences of the 
observer with $u_\mu=-\alpha\delta^t_\mu$. 
For such congruences, we have $u^{t}=\alpha^{-1}$ and 
$u^{k}= -\alpha^{-1}\beta^k~(k=r,~\theta,~\phi)$. 
According with such congruences, the metric can be expressed as 
%
\begin{eqnarray}
ds^2&=&
-\alpha^2 dt^2 
+\frac{1}{\gamma^{rr}}\left(dr+\beta^r dt\right)^2
+\gamma_{\theta\theta}d\theta^2
+\gamma_{\phi\phi}\left[d\phi+\frac{\gamma_{r\phi}}{\gamma_{\phi\phi}}
		\left(dr+\beta^r dt\right)\right]^2, \\
&=&
-\left(\frac{\Sigma}{\Sigma+2mr}\right)dt^2
+\frac{\Sigma(\Sigma+2mr)}{A}
	\left(dr+\frac{2mr}{\Sigma+2mr}dt\right)^2
+\Sigma d\theta^2
+\frac{A\sin^2\theta}{\Sigma}
	\left[
		d\phi-\omega dt-\frac{a}{A}(\Sigma+2mr)dr
	\right]^2,  
\end{eqnarray}
%
where we use $\gamma^{rr}=\gamma_{\phi\phi}/(\gamma_{rr}\gamma_{\phi\phi}-\gamma_{r\phi}^2)$. 
By using the tetrad transformation $x^{\hat{\nu}}=e_\mu^{~~\hat{\nu}}x^\mu$ where the hat denote 
the components measured in the LNRF having the orthonormal tetrad basis, 
the metric can be calculated as  
\begin{eqnarray}
ds^2=
-\left[e_\mu^{~~\hat{t}}dx^\mu\right]^2
+\left[e_\mu^{~~\hat{r}}dx^\mu\right]^2
+\left[e_\mu^{~~\hat{\theta}}dx^\mu\right]^2
+\left[e_\mu^{~~\hat{\phi}}dx^\mu\right]^2.   
\end{eqnarray}
Then, the components of the tetrad $e_\mu^{~~\hat{\nu}}$ connecting between the KSF 
and the LNRF are calculated as, 
%
\begin{eqnarray}
\left(
	\begin{array}{rrrr}
	~
	e_t^{~~\hat{t}}          ~~& 
	e_t^{~~\hat{r}}          ~~& 
	e_t^{~~\hat{\theta}}     ~~& 
	e_t^{~~\hat{\phi}} 	     ~\\
	~
	e_r^{~~\hat{t}}          ~~& 
	e_r^{~~\hat{r}}          ~~& 
	e_r^{~~\hat{\theta}}     ~~& 
	e_r^{~~\hat{\phi}}       ~\\
	~
	e_\theta^{~~\hat{t}}     ~~& 
	e_\theta^{~~\hat{r}}     ~~& 
	e_\theta^{~~\hat{\theta}}~~& 
	e_\theta^{~~\hat{\phi}}  ~\\
	~
	e_\phi^{~~\hat{t}}       ~~& 
	e_\phi^{~~\hat{r}}       ~~& 
	e_\phi^{~~\hat{\theta}}  ~~& 
	e_\phi^{~~\hat{\phi}}    ~\\ 
	\end{array}
\right)
&=&
\left(
	\begin{array}{cccc}
	~
	\alpha 
		~~& \beta^r (\gamma^{rr})^{-1/2}
		~~& 0 
		~~& \beta^r \gamma_{r\phi} (\gamma_{\phi\phi})^{-1/2}
		\\
	~
	0 
		~~& (\gamma^{rr})^{-1/2}  
		~~& 0 
		~~& \gamma_{r\phi} (\gamma_{\phi\phi})^{-1/2}  
		\\
	~
	0 
		~~& 0 
		~~& (\gamma_{\theta\theta})^{1/2}
		~~& 0 
		\\
	~
	0 
		~~& 0 
		~~& 0 
		~~& (\gamma_{\phi\phi})^{1/2}  
		\\
	\end{array}
\right)\\ 
&=&
\left(
	\begin{array}{cccc}
	\displaystyle \left(\frac{\Sigma}{\Sigma+2mr}\right)^{1/2} 
		& \displaystyle \frac{2mr}{A^{1/2}}\left(\frac{\Sigma}{\Sigma+2mr}\right)^{1/2}
		& 0 
		& \displaystyle -\frac{2mra\sin\theta}{(\Sigma A)^{1/2}} \\
	0 
		& \displaystyle \left[\frac{\Sigma(\Sigma+2mr)}{A}\right]^{1/2}  
		& 0 
		& \displaystyle -\frac{a\sin\theta(\Sigma+2mr)}{(A\Sigma)^{1/2}}  \\
	0 
		& 0 
		& \Sigma^{1/2} 
		& 0 \\
	0 
		& 0 
		& 0 
		& \displaystyle \sin\theta\left(\frac{A}{\Sigma}\right)^{1/2}  \\
	\end{array}
\right). 
\end{eqnarray}
By using 
$e^{\mu\hat{\nu}}=g^{\mu\lambda}e_\lambda^{~~\hat{\nu}}$ and 
$e^\mu_{~~\hat{\nu}}=\eta_{\hat{\nu}\hat{\lambda}}e^{\mu\hat{\lambda}}$ 
where $(\eta_{\hat{\mu}\hat{\nu}})$ is the metric of the Lorentz frame, 
i.e. diag$(\eta_{\hat{\mu}\hat{\nu}})=(-1,1,1,1)$ and 
the non-diagonal components of $\eta_{\hat{\mu}\hat{\nu}}$ are null,  
we also have the components for the tetrad $e^\mu_{~~\hat{\nu}}$ calculated as 
%
\begin{eqnarray}
\left(
	\begin{array}{rrrr}
	~
	e^t_{~~\hat{t}}           ~~& 
	e^t_{~~\hat{r}}           ~~& 
	e^t_{~~\hat{\theta}}      ~~& 
	e^t_{~~\hat{\phi}}        ~\\
	~
	e^r_{~~\hat{t}}           ~~& 
	e^r_{~~\hat{r}}           ~~& 
	e^r_{~~\hat{\theta}}      ~~& 
	e^r_{~~\hat{\phi}}        ~\\
	~
	e^\theta_{~~\hat{t}}      ~~& 
	e^\theta_{~~\hat{r}}      ~~& 
	e^\theta_{~~\hat{\theta}} ~~& 
	e^\theta_{~~\hat{\phi}}   ~\\
	~
	e^\phi_{~~\hat{t}}        ~~& 
	e^\phi_{~~\hat{r}}        ~~& 
	e^\phi_{~~\hat{\theta}}   ~~& 
	e^\phi_{~~\hat{\phi}}     ~\\ 
	\end{array}
\right)
&=&
\left(
	\begin{array}{cccc}
	~
	\alpha^{-1} 
		~~& 0
		~~& 0 
		~~& 0
		\\
	~
	-\beta^r \alpha^{-1} 
		~~& (\gamma^{rr})^{1/2}  
		~~& 0 
		~~& 0  
		\\
	~
	0 
		~~& 0 
		~~& (\gamma^{\theta\theta})^{1/2}
		~~& 0 
		\\
	~
	0 
		~~& \gamma^{r\phi} (\gamma^{rr})^{-1/2}
		~~& 0 
		~~& (\gamma_{\phi\phi})^{-1/2}  
		\\
	\end{array}
\right)\\ 
&=&
\left(
	\begin{array}{cccc}
	\displaystyle \left(\frac{\Sigma+2mr}{\Sigma}\right)^{1/2} 
		& 0 
		& 0 
		& 0 \\
	\displaystyle \frac{-2mr}{[\Sigma(\Sigma+2mr)]^{1/2}} 
		& \displaystyle \left[\frac{A}{\Sigma(\Sigma+2mr)}\right]^{1/2} 
		& 0 
		& 0\\
	0 
		& 0 
		& \displaystyle \frac{1}{\Sigma^{1/2}} 
		& 0 \\
	0 
		& \displaystyle a\left(\frac{\Sigma+2mr}{\Sigma A}\right)^{1/2} 
		& 0 
		& \displaystyle \frac{1}{\sin\theta}\left(\frac{\Sigma}{A}\right)^{1/2} \\
	\end{array}
\right). \nonumber\\
\end{eqnarray}
%

%
Since we now consider the accretion flows in the equatorial plane, 
we assume that the physical quantities measured in the LNRF are transformed 
to the physical quantities measured in the FRF by 
two-dimensional Lorentz transformation with the radial velocity $\hat{v}_r$ and 
the azimuthal velocity $\hat{v}_\phi$. 
Here, we assume that the FRF moves with the radial velocity $\hat{v}_r$ and 
the azimuthal velocity $\hat{v}_\phi$ with respect to the LNRF. 
Inversely, the LNRF moves with the radial velocity $-\hat{v}_r$ and 
the azimuthal velocity $-\hat{v}_\phi$ with respect to the FRF. 
The tetrads $e_{\hat{\nu}}^{~(\lambda)}$ and $e^{\hat{\nu}}_{~(\lambda)}$ 
connecting between the LNRF and the FRF 
are two-dimensional Lorentz transformation with the radial velocity $\hat{v}_r$ and 
the azimuthal velocity $\hat{v}_\phi$ measured in the LNRF. 
The transformation matrix $e^{\hat{\nu}}_{~(\lambda)}$ are described as 
\begin{equation}
\left(
	\begin{array}{cccc}
	e^{\hat{t}}_{~(t)} & e^{\hat{t}}_{~(r)} 
		& e^{\hat{t}}_{~(\theta)} & e^{\hat{t}}_{~(\phi)} \\ 
	e^{\hat{r}}_{~(t)} & e^{\hat{r}}_{~(r)} 
		& e^{\hat{r}}_{~(\theta)} & e^{\hat{r}}_{~(\phi)} \\ 
	e^{\hat{\theta}}_{~(t)} & e^{\hat{\theta}}_{~(r)} 
		& e^{\hat{\theta}}_{~(\theta)} & e^{\hat{\theta}}_{~(\phi)} \\ 
	e^{\hat{\phi}}_{~(t)} & e^{\hat{\phi}}_{~(r)} 
		& e^{\hat{\phi}}_{~(\theta)} & e^{\hat{\phi}}_{~(\phi)} \\ 
	\end{array}
\right) 
=
\left(
	\begin{array}{cccc}
	\hat{\gamma}
		& \hat{\gamma}\hat{v}_r
		& 0
		& \hat{\gamma}\hat{v}_\phi \\
	\hat{\gamma}\hat{v}_r
		& \displaystyle 1+\frac{\hat{\gamma}^2\hat{v}_r^2}{1+\hat{\gamma}}
		& 0
		& \displaystyle 
		\frac{\hat{\gamma}^2\hat{v}_r\hat{v}_\phi}{1+\hat{\gamma}} \\
	0
		& 0
		& 1
		& 0 \\
	\hat{\gamma}\hat{v}_\phi
		& \displaystyle \frac{\hat{\gamma}^2\hat{v}_r
			\hat{v}_\phi}{1+\hat{\gamma}}
		& 0
		& \displaystyle 
			1+\frac{\hat{\gamma}^2 \hat{v}_\phi^2}{1+\hat{\gamma}} 
	\end{array}
\right)
\end{equation}
where $\hat\gamma=(1-\hat{v}_r^2-\hat{v}_\phi^2)^{-1/2}$. 
Since both the LNRF and the FRF are orthonormal, 
the transformation matrix $e_{\hat{\nu}}^{~(\lambda)}$ are calculated 
as 
\begin{equation}
\left(
	\begin{array}{cccc}
	e_{\hat{t}}^{~(t)} & e_{\hat{t}}^{~(r)} 
		& e_{\hat{t}}^{~(\theta)} & e_{\hat{t}}^{~(\phi)} \\ 
	e_{\hat{r}}^{~(t)} & e_{\hat{r}}^{~(r)} 
		& e_{\hat{r}}^{~(\theta)} & e_{\hat{r}}^{~(\phi)} \\ 
	e_{\hat{\theta}}^{~(t)} & e_{\hat{\theta}}^{~(r)} 
		& e_{\hat{\theta}}^{~(\theta)} & e_{\hat{\theta}}^{~(\phi)} \\ 
	e_{\hat{\phi}}^{~(t)} & e_{\hat{\phi}}^{~(r)} 
		& e_{\hat{\phi}}^{~(\theta)} & e_{\hat{\phi}}^{~(\phi)} \\ 
	\end{array}
\right) 
=
\left(
	\begin{array}{cccc}
	e^{\hat{t}}_{~(t)} & -e^{\hat{t}}_{~(r)} 
		& -e^{\hat{t}}_{~(\theta)} & -e^{\hat{t}}_{~(\phi)} \\ 
	-e^{\hat{r}}_{~(t)} & e^{\hat{r}}_{~(r)} 
		& e^{\hat{r}}_{~(\theta)} & e^{\hat{r}}_{~(\phi)} \\ 
	-e^{\hat{\theta}}_{~(t)} & e^{\hat{\theta}}_{~(r)} 
		& e^{\hat{\theta}}_{~(\theta)} & e^{\hat{\theta}}_{~(\phi)} \\ 
	-e^{\hat{\phi}}_{~(t)} & e^{\hat{\phi}}_{~(r)} 
		& e^{\hat{\phi}}_{~(\theta)} & e^{\hat{\phi}}_{~(\phi)} \\ 
	\end{array}
\right). 
\end{equation}
Since now we have both the tetrads between the KSF and the LNRF and the tetrad between the LNRF and 
the KSF, the tetrad connecting the KSF and the FRF, 
e.g. $e^\mu_{~~(\nu)}$ and $e_\mu^{~~(\nu)}$, are calculated as 
\begin{equation}
e^\mu_{~~(\nu)}=e^{\mu}_{~~\hat{\lambda}}e^{\hat{\lambda}}_{~~(\nu)},~~~~
e_\mu^{~~(\nu)}=e_{\mu}^{~~\hat{\lambda}}e_{\hat{\lambda}}^{~~(\nu)}. 
\end{equation}
By using these tetrad, we can transform the physical quantities between the KSF and the FRF as 
%
\begin{eqnarray}
&&u^\mu=e^\mu_{~~(\nu)}u^{(\nu)},~~~u_\mu=e_\mu^{~~(\nu)}u_{(\nu)}, \\
&&t^{\mu\nu}=e^\mu_{~~(\lambda)}e^\nu_{~~(\xi)}t^{(\lambda)(\xi)},~~~ 
t_{\mu\nu}=e_\mu^{~~(\lambda)}e_\nu^{~~(\xi)}t_{(\lambda)(\xi)},~~~
t^{\mu}_{~~\nu}=e^{\mu(\lambda)}e_\nu^{~~(\xi)}t_{(\lambda)(\xi)},~~~
\end{eqnarray}
%
and so on. 
Lowering and raising the index $\mu$ of the tetrads are done by $g_{\mu\nu}$ and $g^{\mu\nu}$. 
On the other hand, 
lowering and raising the index $\mu$ of the tetrads are done by $\eta_{(\mu)(\nu)}$ 
and $\eta^{(\mu)(\nu)}$ which are the metric of the Lorentz frame. 

\section{Velocity Fields}
\label{app:velocity_field}
By using the tetrads derived in Appendix \ref{app:tetrads}, in this appendix, we first derive 
the velocity fields $u^\mu$ and $u_\mu$ in the KSF described by the radial velocity $\hat{v}_r$ and 
the azimuthal velocity $\hat{v}_\phi$ of the FRF with respect to the LNRF. 
Since $u^{(t)}=-u_{(t)}=1$ and $u^{(k)}=u_{(k)}=0$ $(k=r,~\theta,~\phi)$, the four velocity is 
calculated as 
$u^\mu = e^\mu_{~~(\nu)} u^{(\nu)} = e^\mu_{~~(t)} u^{(t)} = e^\mu_{~~(t)}
= e^\mu_{~~\hat{\lambda}}e^{\hat{\lambda}}_{~~(t)}$ 
and 
$u_\mu = e_\mu^{~~(\nu)} u_{(\nu)} = e_\mu^{~~(t)} u_{(t)} = -e_\mu^{~~(t)}
=-e_\mu^{~~\hat{\lambda}}e_{\hat{\lambda}}^{~~(t)}$. 
Then, we have 
%
\begin{equation}
\left(
	\begin{array}{c}
		u^t \\
		u^r \\
		u^\theta \\
		u^\phi
	\end{array}
\right)
= 
\left(
	\begin{array}{c}
		\alpha^{-1} \hat{\gamma}\\
		-\beta^r \alpha^{-1} \hat{\gamma}  
		+ (\gamma^{rr})^{1/2} \hat{\gamma} \hat{v}_r 
		\\
		0 \\
		\gamma^{r\phi} (\gamma^{rr})^{-1/2} \hat{\gamma} \hat{v}_r
			+ (\gamma_{\phi\phi})^{-1/2} \hat{\gamma} \hat{v}_\phi
	\end{array}
\right),
\end{equation}
and 
\begin{equation}
\left(
	\begin{array}{c}
		u_t \\
		u_r \\
		u_\theta \\
		u_\phi
	\end{array}
\right)
= 
\left(
	\begin{array}{c}
		-\alpha\hat{\gamma} 
			+\beta^r (\gamma^{rr})^{-1/2} \hat{\gamma} \hat{v}_r 
			+\beta^r \gamma_{r\phi} (\gamma_{\phi\phi})^{-1/2} 
			\hat{\gamma} \hat{v}_\phi
		\\
		(\gamma^{rr})^{-1/2} \hat{\gamma} \hat{v}_r 
		+\gamma_{r\phi} (\gamma_{\phi\phi})^{-1/2} \hat{\gamma} 
		\hat{v}_\phi
		\\
		0 \\
		(\gamma_{\phi\phi})^{1/2} \hat{\gamma} \hat{v}_\phi
	\end{array}
\right). 
\end{equation}
%
From $u^t$, $u^r$ and $u_\phi(=\ell)$, we can derive Eq. (\ref{eq:gam_vr_vphi}) as 
\begin{equation}
\hat{v}_r = \frac{u^r+\beta^r u^t}{\hat{\gamma}(\gamma_{rr})^{1/2}}
	=\frac{\Sigma+2mr}{u^t A^{1/2}}\left(u^r+\frac{2mr}{\Sigma+2mr}u^t\right),~~~~
\hat{v}_\phi = \frac{\ell}{\hat{\gamma}(\gamma_{\phi\phi})^{1/2}}
	=\frac{\ell}{u^t}\left(\frac{\Sigma+2mr}{A}\right)^{1/2},
\end{equation} 
where we use $\theta=\pi/2$.

\section{Shear Rate}
\label{app:shear}
In this appendix, we calculate the coefficients $\sigma_r$, $\sigma_u$ and $\sigma_\ell$ which 
gives the shear rate $\sigma(\equiv\sigma_{(r)(\phi)})$ measured in the FRF by 
\begin{eqnarray}
\sigma=\sigma_r+\sigma_u\frac{du^r}{dr}+\sigma_\ell\frac{d\ell}{dr}. \nonumber
\end{eqnarray}
The shear rate $\sigma_{(r)(\phi)}$ in the FRF is calculated from the shear rate 
$\sigma_{\mu\nu}$ in the KSF as 
%
$\sigma_{(r)(\phi)}=e^\mu_{~(r)}e^\nu_{~(\phi)}\sigma_{\mu\nu}$.  
%
Here, non-zero components of the tetrads are 
$e^\mu_{~(r)}$ ($\mu=t,~r,~\phi$) and $e^\nu_{~(\phi)}$ ($\nu=t,~r,~\phi$) which can be 
calculated by 
the tetrad connecting the KSF and the LNRF, $e^{\mu}_{~\hat{\lambda}}$, and 
the tetrad connecting the LNRF and the FRF, $e^{\hat{\lambda}}_{~(\alpha)}$ ($\alpha=r,~\phi$) 
given in Appendix \ref{app:tetrads}.  
Since $\sigma_{\mu\nu}=\sigma_{\nu\mu}$, we need to calculate 
six components of the shear rate, i.e. 
$\sigma_{tt}$, $\sigma_{tr}$, $\sigma_{t\phi}$, 
$\sigma_{rr}$, $\sigma_{r\phi}$ and $\sigma_{\phi\phi}$.  
The shear rate $\sigma_{\mu\nu}$ 
is calculated as the traceless part of the deformation tensor which is 
calculated as 
%
$
\sigma_{\mu\nu}=
	(u_{\mu;\alpha}h^\alpha_\nu +u_{\nu;\alpha}h^\alpha_\mu)/2
	-\Theta h_{\mu\nu}/3. 
$
%
%
We give the covariant derivative for $u_\mu$ and the four acceleration $a_\mu$ 
which are directly used for the calculations of the shear rate $\sigma_{\mu\nu}$. 
The non-zero components of $u_{\mu;nu}$ are calculated as 
%
\begin{eqnarray}
&&
u_{t;t}   =\frac{1}{2}g_{tt,r}u^r,~~~~
u_{r;t}   =-\frac{1}{2}(g_{tt,r}+\Omega g_{t\phi,r})u^t,~~~~
u_{\phi;t}=\frac{1}{2}g_{t\phi,r}u^r,~~~~
u_{t;r}   =-\frac{d\mathcal{E}}{dr}-\frac{1}{2}(g_{tt,r}+\Omega g_{t\phi,r})u^t,\nonumber\\
&&
u_{r;r}   =\frac{du_r}{dr}-\frac{1}{2}g_{rr,r}u^r-(g_{tr,r}+\Omega g_{r\phi,r})u^t,~~~~
u_{\phi;r}=\frac{d\ell}{dr}-\frac{1}{2}(g_{t\phi,r}+\Omega g_{\phi\phi,r})u^t,~~~~
u_{\theta;\theta}=\frac{1}{2}g_{\theta\theta,r}u^r,\nonumber\\
&&
u_{t;\phi}   =\frac{1}{2}g_{t\phi,r}u^r,~~~~
u_{r;\phi}   =-\frac{1}{2}(g_{t\phi,r}+\Omega g_{\phi\phi,r})u^t,~~~~
u_{\phi;\phi}=\frac{1}{2}g_{\phi\phi,r}u^r.  
\label{eq:u_mu;nu}
\end{eqnarray}
%
The covariant components of the four acceleration, $a_\mu=u_{\mu;\nu}u^\nu$, 
are calculated as 
%
\begin{equation}
a_t=-u^r\frac{d\mathcal{E}}{dr},~~~
a_r=u^r\frac{du_r}{dr} 
	-\frac{1}{2}g_{rr,r}(u^r)^2
	-(g_{tr,r}+g_{r\phi,r}\Omega)u^t u^r
	-\frac{1}{2}(u^t)^2g_{\phi\phi,r}(\Omega-\Omega_K^+)(\Omega-\Omega_K^-),~~~
a_\theta=0,~~~
a_\phi=u^r\frac{d\ell}{dr}.
\end{equation}
%

%
We next calculate the expansion $\Theta=u^\gamma_{~;\gamma}$. 
The covariant derivatives $u^t_{~;t}$, $u^r_{~;r}$, $u^\theta_{~;\theta}$ and 
$u^\phi_{~;\phi}$ are calculated as 
%
\begin{eqnarray}
u^t_{~;t}&=& 
	\frac{1}{2}(g^{tt}g_{tt,r}+g^{t\phi}g_{t\phi,r})u^r
	-\frac{1}{2}g^{tr}(g_{tt,r}+g_{t\phi,r}\Omega)u^t
	,\\
u^r_{~;r}&=&
	\frac{du^r}{dr}+\frac{1}{2}g^{rr}g_{rr,r}u^r
	+(g^{tr}g_{tr,r}+g^{r\phi}g_{r\phi,r})u^r
	+\frac{1}{2}\left[
		g^{tr}(g_{tt,r}+g_{t\phi,r}\Omega)
		+g^{r\phi}(g_{t\phi,r}+g_{\phi\phi,r}\Omega)
	\right]u^t
	,\\
u^\theta_{~;\theta}&=&
	\frac{1}{2}g^{\theta\theta}g_{\theta\theta,r}u^r
	,\\
u^\phi_{~;\phi}&=&
	\frac{1}{2}(g^{t\phi}g_{t\phi,r}+g^{\phi\phi}g_{\phi\phi,r})u^r
	-\frac{1}{2}g^{r\phi}(g_{t\phi,r}+g_{\phi\phi,r}\Omega)u^t
	. 
\end{eqnarray}
%
Then, the expansion $\Theta$ is calculated by the form 
\begin{equation}
\Theta=\frac{du^r}{dr}+\Theta_r, 
\end{equation}
where $\Theta_r$ is defined as 
\begin{equation}
\Theta_r=
	\frac{u^r}{2}\left(
		g^{tt}g_{tt,r}+g^{rr}g_{rr,r}+g^{\theta\theta}g_{\theta\theta,r}
		+g^{\phi\phi}g_{\phi\phi,r}	
		+2g^{tr}g_{tr,r}+2g^{t\phi}g_{t\phi,r}+2g^{r\phi}g_{r\phi,r}
	\right)
\end{equation}

%
We finally calculate the shear rate $\sigma_{(r)(\phi)}$ which can be calculated as 
$\sigma_{(r)(\phi)}=\sigma_r+\sigma_u(du^r/dr)+\sigma_\ell(d\ell/dr)$. 
By using the tetrads are given in Appendix \ref{app:tetrads}, the shear rate $\sigma_{(r)(\phi)}$ 
measured in the FRF is calculated 
by the shear rate $\sigma_{\mu\nu}$ measured in KSF as 
\begin{eqnarray}
\sigma_{(r)(\phi)}
=e^\mu_{~(r)}e^\nu_{~(\phi)}\sigma_{\mu\nu}
=\frac{1}{2}(u_{\mu;\nu}+u_{\nu;\mu})e^\mu_{~(r)}e^\nu_{~(\phi)}
	-\frac{\Theta}{3}e_\mu^{~(r)}e^\mu_{~(\phi)}. 
\label{eq:sigmaFRF}
\end{eqnarray} 
By substituting $u_{\mu;nu}$ given by Eq. (\ref{eq:u_mu;nu}) into Eq. (\ref{eq:sigmaFRF}), 
we can analytically calculate the coefficients $\sigma_r$, $\sigma_u$ and $\sigma_\ell$. 
%

\section{Calculations for $\ell_*$}
\label{app:ell*}
\subsection{Derivation of $\ell_*$ in Kerr-Schild coordinate}
The explicit form for $\ell_*$ is obtained by the direct calculation of 
$\ell_*^2=r^2(\Gamma^\mu_{\theta\nu}u_\mu u^\nu)_1$ where $(\Gamma^\mu_{\theta\nu}u_\mu u^\nu)_1$ 
is calculated from $\Gamma^\mu_{\theta\nu}u_\mu u^\nu$ until the order of $\cos\theta$. 
$2\Gamma^\mu_{\theta\nu}u_\mu u^\nu$ is calculated as 
\begin{equation}
2\Gamma^\mu_{\theta\nu}u_\mu u^\nu
=g_{tt,\theta}(u^t)^2+g_{rr,\theta}(u^r)^2+g_{\phi,\theta}(u^\phi)^2+
+2g_{tr,\theta}u^t u^r+2g_{t\phi,\theta}u^t u^\phi+2g_{r\phi,\theta}u^r u^\phi, 
\label{eq:2Gamma}
\end{equation}
where the derivatives of the metric components with respect to $\theta$ are calculated 
until the order of $\cos\theta$ around the equatorial plane $\theta=\pi/2$ as 
\begin{eqnarray}
&&g_{tt,\theta}=g_{tr,\theta}=g_{rr,\theta}=\frac{2ma^2}{r^3}(2\cos\theta),~~~~
g_{t\phi,\theta}=-\frac{2ma}{r}\left(1+\frac{a^2}{r^2}\right)(2\cos\theta),\nonumber\\
&&
g_{r\phi,\theta}=-a\left[1+\frac{2m}{r}\left(1+\frac{a^2}{r^2}\right)\right](2\cos\theta),~~~~
g_{\phi\phi,\theta}=r^2\left[1+\frac{a^2}{r^2}
+\frac{2ma^2}{r^3}\left(2+\frac{a^2}{r^2}\right)\right](2\cos\theta),\label{eq:gtheta}
\end{eqnarray}
and $u^t$, $u^r$ and $u^\phi$ are expressed by $\mathcal{E}$, $\ell$ and $u_r$ as 
\begin{equation}
u^t=\mathcal{E}\left(1+\frac{2m}{r}\right)+\frac{2m}{r}u_r,~~~~
u^r=-\frac{2m}{r}\mathcal{E}+\frac{\Delta}{r^2}u_r+\frac{a}{r^2}\ell,~~~~
u^\phi=\frac{1}{r^2}(\ell+au_r). \label{eq:uturuphi}
\end{equation}
By substituting $u^t$, $u^r$ and $u^\phi$ given by Eq. (\ref{eq:uturuphi}), we obtain 
the equation including the term with $(u_r)^2$ which is calculated from $u^\mu u_\mu=-1$ as  
\begin{equation}
\frac{\Delta}{r^2}(u^r)^2=
	\mathcal{E}\left(1+\frac{2m}{r}\right)
	-\frac{\ell^2}{r^2}
	-1
	+2u_r\left(\frac{2m}{r}\mathcal{E}-\frac{a}{r^2}\ell\right). \label{eq:u_r^2}
\end{equation}
From Eq. (\ref{eq:2Gamma}) with Eqs. (\ref{eq:gtheta}), (\ref{eq:uturuphi}) and (\ref{eq:u_r^2}), 
we can finally obtain $(\Gamma^\mu_{\theta\nu}u_\mu u^\nu)_1$ as 
\begin{equation}
(\Gamma^\mu_{\theta\nu}u_\mu u^\nu)_1=\frac{\ell^2-a^2(\mathcal{E}^2-1)}{r^2}, \label{eq:Gamma}
\end{equation}
where the coefficients of $u_r$ become null. 
As shown in \cite{alp97}, in the Boyer-Lindquist coordinate we obtain the same expression as 
Eq. (\ref{eq:Gamma}). 
This is because the specific energy $\mathcal{E}$ and the angular momentum $\ell$ have 
the same expression for both coordinate, i.e. $\mathcal{E}_{\rm BL}=\mathcal{E}_{\rm KS}$ and 
$\ell_{\rm BL}=\ell_{\rm KS}$, 
as shown by the transformations of four velocities given by Eq. (\ref{eq:uBLKS2}). 
Here, "BL" and "KS" denote the physical quantities calculated by using the Boyer-Lindquist 
coordinate and the Kerr-Schild coordinate, respectively.

\subsection{Derivative of $\ell_*$ with respect to $r$}
\label{app:dell*/dr}
We express the derivative $d\ell_*/dr$ by the combination of $du^r/dr$ and $d\ell/dr$ as 
\begin{equation}
\frac{d\ell_*}{dr}=\ell_*^r+\ell_*^u\frac{du^r}{dr}+\ell_*^\ell\frac{d\ell}{dr}, 
\end{equation}
where $\ell_*^r$, $\ell_*^u$ and $\ell_*^\ell$ are calculated as 
\begin{equation}
\ell_*^r=-\frac{a^2\mathcal{E}}{\ell_*}\mathcal{E}_r,~~~~
\ell_*^u=-\frac{a^2\mathcal{E}}{\ell_*}\mathcal{E}_u,~~~~
\ell_*^\ell=-\frac{a^2\mathcal{E}}{\ell_*}\mathcal{E}_\ell+\frac{\ell}{\ell_*}.  
\end{equation}
Here, $\mathcal{E}_r$, $\mathcal{E}_u$ and $\mathcal{E}_\ell$ are defined to have the relation 
$d\mathcal{E}/dr=\mathcal{E}_r+\mathcal{E}_u (du^r/dr)+\mathcal{E}_\ell(\ell/dr)$ and are 
calculated fully analytically or numerically. 
%



\bsp

\label{lastpage}

\end{document}